\documentclass[a4paper,fleqn,usenatbib]{mnras}

\usepackage[T1]{fontenc}
\usepackage{ae,aecompl,color,pdflscape,here}

\usepackage{graphicx}	% Including figure files
\usepackage{amsmath}	% Advanced maths commands
\usepackage{amssymb}	% Extra maths symbols
\hypersetup{draft}
\title[A {\it Gaia} view of Cyg OB2 and Car OB1]{A {\it Gaia} view of the two OB associations Cygnus OB2 and 
Carina OB1: The signature of their formation process}

\author[B. Lim et al.]{Beomdu Lim$^{1,2,3}$\thanks{Corresponding author, E-mail:blim@khu.ac.kr}, Ya\"el 
Naz\'e$^2$\thanks{Research Associate FRS-FNRS (Belgium)}, Eric Gosset$^2$\thanks{Research Director FRS-FNRS (Belgium)}, and Gregor Rauw$^2$ \\
$^1$School of Space Research, Kyung Hee University, 1732 Deogyeong-daero, Giheung-gu, 
Yongin-si, Gyeonggi-do 17104, \\ 
Republic of Korea\\
$^2$Space sciences, Technologies and Astrophysics Research Institute, 
                  Universit\'e de Li\`ege, Quartier Agora, All\'ee du 6 Ao\^ut 19c, B\^at. \\ 
B5c, 4000, Li\`ege, Belgium\\
$^3$Department of Astronomy and Space Science, Sejong University, 209 Neungdong-ro, 
              Gwangjin-gu, Seoul 05006, Republic of Korea\\
}

\date{Accepted XXX. Received YYY; in original form ZZZ}

\pubyear{2019}

\begin{document}
\label{firstpage}
\pagerange{\pageref{firstpage}--\pageref{lastpage}}
\maketitle

\begin{abstract}
OB associations are the prime star forming sites in galaxies. However the detailed 
formation process of such stellar systems still remains a mystery. In this 
context, identifying the presence of substructures may help tracing the footprints of 
their formation process. Here, we present a kinematic study of the two massive 
OB associations Cygnus OB2 and Carina OB1 using the precise 
astrometry from the {\it Gaia} Data Release 2 and radial velocities. 
From the parallaxes of stars, these OB associations are confirmed to be genuine 
stellar systems. Both Cygnus OB2 and Carina OB1 are composed of a few dense clusters 
and a halo which have different kinematic properties: the clusters 
occupy regions of 5--8 parsecs in diameter and display small dispersions in proper 
motion, while the halos spread over tens of parsecs with a 2-3 times larger 
dispersions in proper motion. This is reminiscent of the 
so-called ``line width-size" relation of molecular clouds related to turbulence. 
Considering that the kinematics and structural features were inherited from 
those of their natal clouds would then imply that the formation of OB associations 
may result from structure formation driven by supersonic turbulence, rather 
than from the dynamical evolution of individual embedded clusters.
\end{abstract}


%%%%%%%%%%%%%%%%%%%%%%%%%%%%%%%%%%%%%%%%%%%%%%%%%%

%%%%%%%%%%%%%%%%% BODY OF PAPER %%%%%%%%%%%%%%%%%%

\section{Introduction}
OB associations are huge stellar systems incubating loose groups of O- and 
B-type stars spread over tens of parsecs \citep{A47}. These stellar systems are 
the prevailing star forming sites in external galaxies as well as in the 
Galaxy \citep{RW93,BKS96,PGFP01,GHV09}, and they are considered as 
the birth places of field stars \citep{MS78,BPS07}. In addition, OB associations 
play a crucial role in the chemical evolution of host galaxies as massive 
stars in those associations produce heavy elements through supernova explosions. 
Despite their importance, our knowledge of such objects remains incomplete, 
in particular the details of their formation process are still not fully established. 
There are two main models concerning the origin of OB associations, 1) the 
expansion of embedded clusters and 2) the formation of unbound stellar 
groups in-situ.

The majority of the stars in star forming regions are thought to form in clusters 
\citep{LL03,PCA03,KAG08}, but only less than 10 per cent of these clusters can 
remain bound, according to comparison of the observed number of clusters 
with the predictions of a model for a constant cluster formation rate \citep{LL03}. 
Most cluster members are then scattered out after gas expulsion \citep{LMD84,
KAH01,BK07}. The role of rapid gas expulsion in disruption of star clusters was 
studied by several groups \citep{T78,H80,GB06}. This dynamical evolution 
leads to the formation of unbound OB associations. 

According to the second model, the origin of OB associations can be explained 
by star formation taking place in hierarchical substructures of molecular clouds. 
Shocks by turbulent flows can create a network of a number of substructures 
inside a molecular cloud \citep{L81,PJGN01,E02}. Both bound clusters and 
distributed population of stars form along these substructures with different 
sizes and densities. Bound clusters preferentially form in high-density 
regions because such regions have high-star formation efficiency, while low-density 
regions with low-star formation efficiency form distributed stellar populations 
\citep{BSCB11,K12}. Gas pressure is also a considerable factor to form either 
bound clusters or sparse groups of stars \citep{E08}. 

\begin{figure*}
   \centering
  \includegraphics[width=5.3cm]{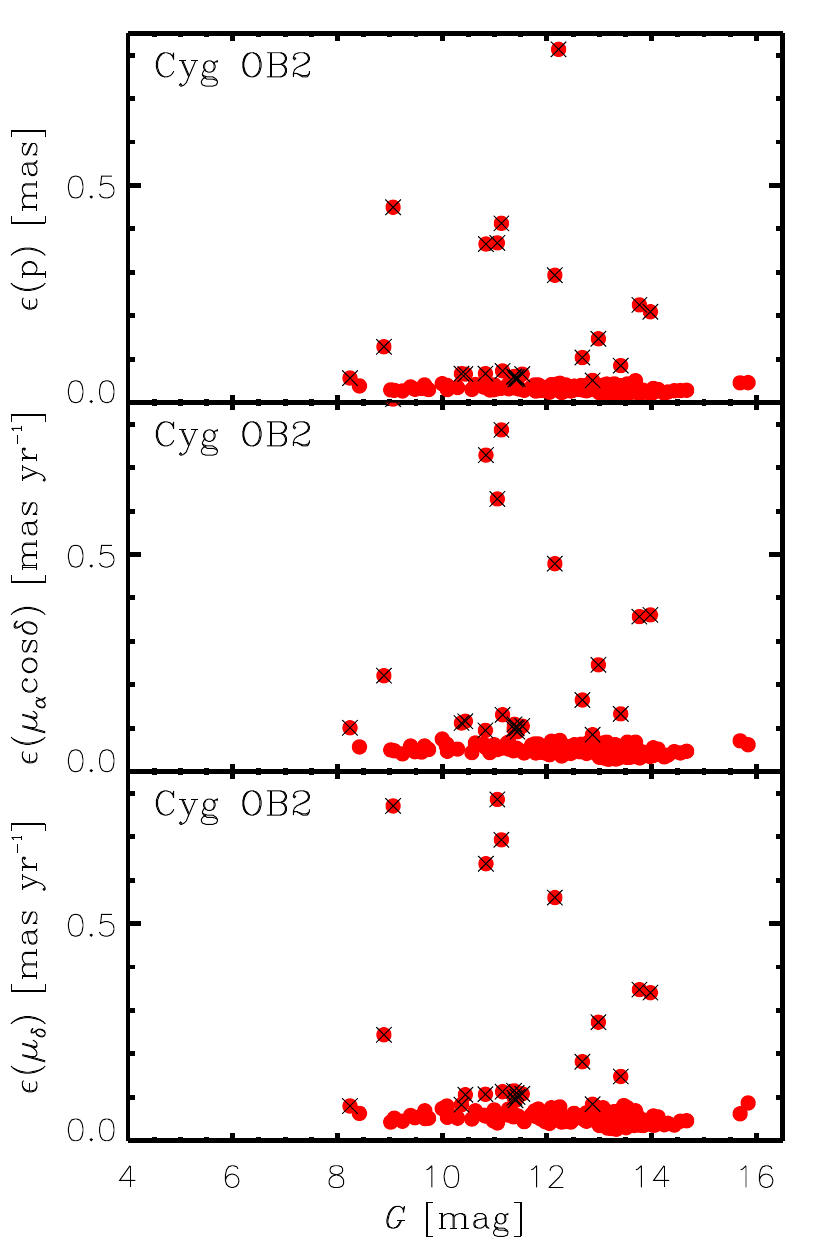}  \includegraphics[width=5.3cm]{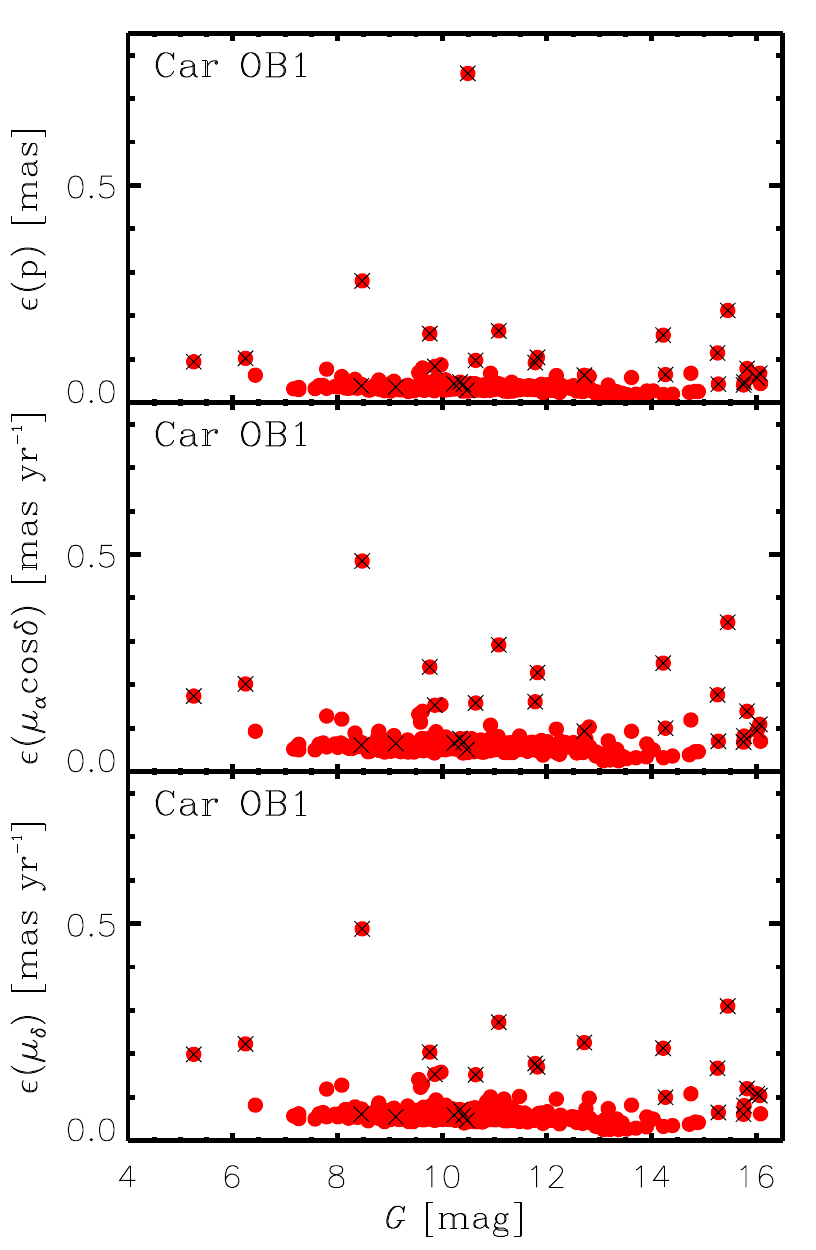}\includegraphics[width=5.3cm]{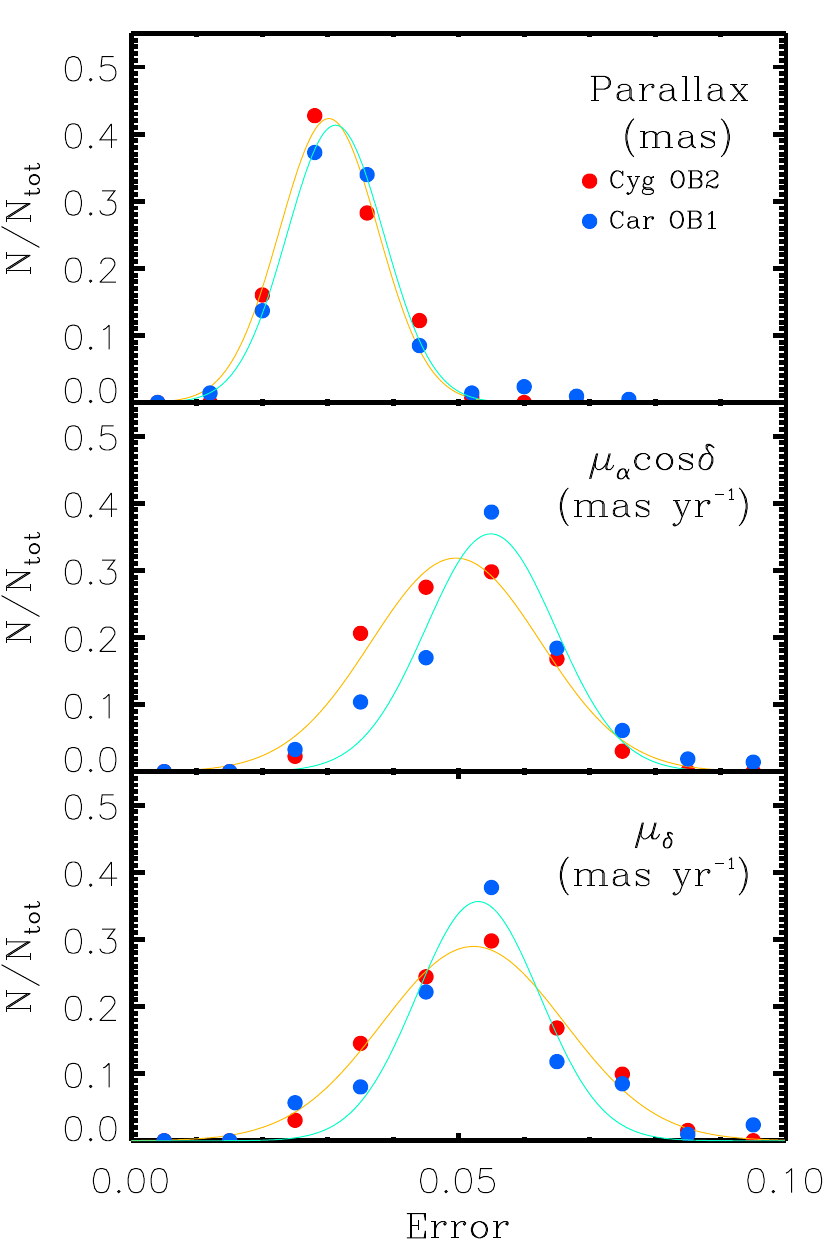}
      \caption{Error distributions of parallaxes and PMs with respect to the $G$ band 
magnitude for stars in Cyg OB2 and Car OB1. In the left and middle panels, crosses 
represent either the data with errors larger than three times the standard deviation 
from the mean of errors in parallax and PM or those with parallaxes smaller than 
five times the associated parallax errors. These data were not used in further 
analysis. The right-hand panel displays the normalised number distributions of errors 
in parallax and PM that we used in this paper. Red and blue dots represent the data for 
Cyg OB2 and Car OB1, respectively. }
         \label{fig1}
 \end{figure*}
 
A number of observational studies have attempted to understand the 
dynamical evolution of OB associations and eventually their formation process \citep{WBD16,MD17,CJW18,KCS18,WK18,WM18,KHS19}. Many OB associations 
seem to be unbound, given that their total stellar masses are smaller than 
virial masses \citep{MD17}. The Orion Molecular Cloud Complex hosts about 
190 subgroups of stars, which are associated with five main components 
composing this complex \citep{KCS18}. One of them, the gas-deficient component 
Orion D (see \citealt{KCS18} for details), exhibits a pattern of expansion. 
\citet{CJW18} also found non-isotropic expansion of stellar subgroups 
in the direction of the Vela-Puppis region. The ages of these subgroups exceed 
the dissipation timescale (10 Myr) of molecular clouds around open clusters 
\citep{LBT89}. There may be little molecular gas around the subgroups. The 
origin of gas-deficient OB associations could be explained by the model 
of expansion of embedded clusters after rapid gas expulsion. 

On the other hand, \citet{WM18} could not find evidence for global expansion 
in the Scorpius-Centaurus OB association, but they found that this association 
is highly substructured. \citet{WK18} investigated 18 nearby OB associations 
using the Tycho-{\it Gaia} Astrometric Solution data \citep{MLH15}. They also could 
not find any signature of global expansion from a single cluster or multiple 
clusters. \citet{CJW18} claimed that the anisotropic expansion and the complicated star 
formation history among the unbound subgroups of stars in the Vela-Puppis 
region may be the footprints of structure formation inside molecular clouds 
driven by turbulent flows. 

Up to today, a number of previous studies have reported different dynamical 
status for various OB associations. Hence, more conclusive evidence for each 
theoretical model is still required to find a suitable explanation for their origin. The 
recent high-precision astrometry from the {\it Gaia} mission \citep{gaia16} 
opens a new window for addressing this issue. In this paper, we use the recent 
astrometric data from the Gaia Data Release 2 (DR2; \citealt{gaia18}) to test the 
two proposed models for the two massive OB associations Cygnus OB2 (Cyg OB2) 
and Carina OB1 (Car OB1). 

Cyg OB2 is the most massive nearby OB association. This association contains 
an abundant population of O- and B-type stars \citep[etc.]{JM54,MT91,CPR02,WHL02,
H03,KKK07} with a total mass reaching about 2--10 $\times \ 10^4 M_{\sun}$ 
\citep{K00,WDDV10,WDM15}. The distance to Cyg OB2 was thought to be about 
$d = 1.4$ kpc \citep{H03,RBS12}, but the association is severely obscured by 
dust lanes, leading to high levels of extinction ($A_V = 4$ -- 20 mag, 
\citealt{K00,WDM15}). Extensive studies of this association showed that Cyg OB2 is 
gravitationally unbound and highly substructured with different kinematics 
\citep{WPGD14,WBD16}. However, no evidence for an overall expanding motion 
of stars was found. Accordingly, \citet{WPGD14,WBD16} claimed that Cyg OB2 
has not originated from the expansion of a single star cluster. \citet{BWHDL19} explored the 
internal substructures of these associations using the parallaxes 
from {\it Gaia} DR2 and found that two stellar groups are lying 
along the line-of-sight. \citet{KKK07} measured the radial velocities (RVs) of 
OB stars in this association, but they could not find any spatial variation of RVs.

Car OB1 association is one of the most active star forming complexes in the 
Carina-Sagittarius spiral arm. This association is supposed to be located 
at 2.2 -- 2.9 kpc \citep{AH93,S06,HSB12}. It contains a large number of 
O- and B-type stars \citep{W73,W95,WHL02,LM82,MGL88,MJ93}, and its total 
mass exceeds $2 \ \times \ 10^4 M_{\sun}$ \citep{PRK11}. Several star clusters, such as 
Trumpler 14 (Tr14), 15 (Tr15), 16 (Tr16), Bochum 10, 11, Collinder 232, 
and 228, as well as halo populations are distributed over this complex 
\citep{FGT11}. However, it is still uncertain whether or not these clusters are 
part of the same association. Tr14, 16, and Collinder 232 
have almost the same mean proper motions (PMs) within the measurement 
errors \citep{CMD93}. In addition, the RVs of stars in the star clusters of 
Car OB1 reveal a single Gaussian distribution although there is a slight 
difference in RVs between Tr14 and Tr16 \citep{KS18}. Thus, these clusters 
are believed to be at the same distance because of their similarities in kinematics. 

Since these associations are young (4--5 Myr for Cyg OB2 -- \citealt{WDM15}; 
1--3 Myr for Car OB1 -- \citealt{HSB12}), the signatures of their formation 
process may still remain detectable in the kinematics as well as in the internal 
structure. To find evidence of the formation process of OB associations, 
we probe the spatial distribution and kinematics of high-mass stars in 
Cyg OB2 and Car OB1. The data we used are described in Section 2. The 
internal structure and kinematics of Cyg OB2 and Car OB1 are explored in 
Sections 3 and 4, respectively. We discuss their formation 
process in Section 5. Finally, our results are summarised in Section 6.

\section{Data} 
The census of high-mass stars (O-, B-type stars, and Wolf-Rayet stars) in Cyg OB2 
and Car OB1 associations is more complete than those of the low-mass star population. 
Such high-mass stars are brighter than $G \sim 16$ mag, and the {\it Gaia} DR2 
provides very high precision astrometric data in that magnitude range \citep{gaia18}.
Therefore, we considered only the high-mass star population of the associations 
in this study. For Cyg OB2, a list of high-mass stars compiled by \citet{WDM15} was 
used. This catalogue contains the spectral types, the stellar parameters, and the photometric 
data of 167 stars earlier than B5 taken from a number of references. For Car OB1, 
\citet{NBO11} studied 200 OB stars [70 O-type and 130 B-type, compiled from 
\citet{S09}]. In addition, we added 106 OB stars from recent photometric or spectroscopic 
studies \citep{HSB12,SMM14,AHPM16,DKJ17,HMP18}. A total of 306 high-mass stars in 
Car OB1 were included in the final list.

\begin{figure*}
   \centering
\includegraphics[width=14cm]{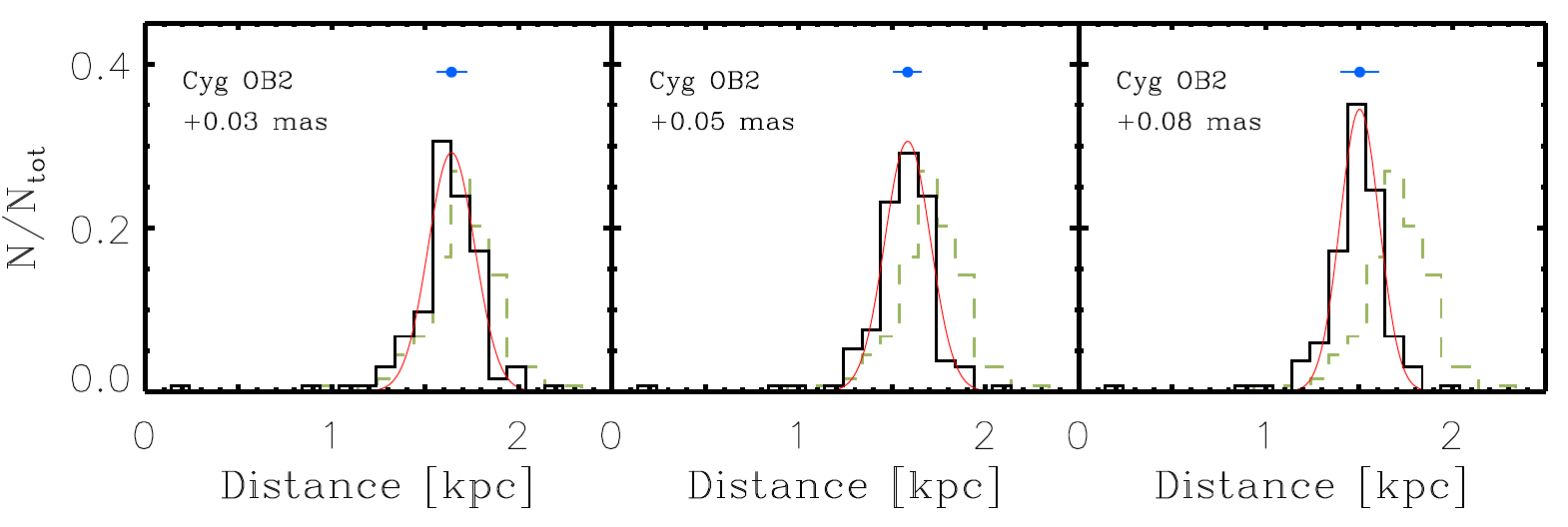}  
\includegraphics[width=14cm]{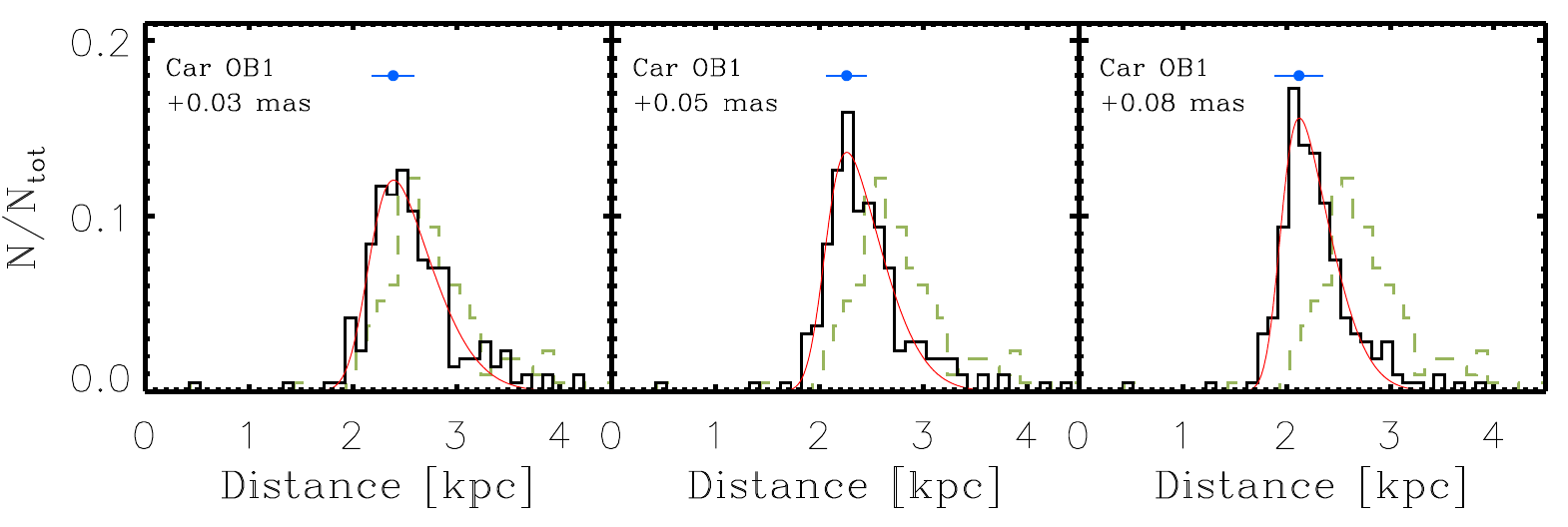}
      \caption{Distance distributions of stars in the direction of Cyg OB2 (upper) 
and Car OB1 (lower). The zero point offsets of 0.03, 0.05, and 0.08 mas were 
applied to the parallaxes of {\it Gaia} DR2 \citep{LHB18,ST18,ZPHS19}. Histograms 
(light green) outlined by light green dashed lines exhibit the distributions of distances 
without correction for the zero point offsets. All the histograms were obtained with a 
bin size of about 0.1 kpc. The error bar at the top of each panel represents the mean 
of errors in distance. In order to find peak distances, Gaussian and skewed Gaussian 
profiles were fitted to the histograms for Cyg OB2 and Car OB 1, respectively (red 
solid lines). In this paper, a mean value obtained from the peak distances (and 
indicated by the blue dot) was adopted as the distance to a given association. }
         \label{fig2}
 \end{figure*}

The parallaxes and PMs of these stars in these associations were taken 
from {\it Gaia} DR 2 \citep{gaia18}. The counterparts of those OB stars 
were searched for within $2\farcs0$ in the {\it Gaia} catalogue. We 
found 165 and 304 counterparts for Cyg OB2 and Car OB1, 
respectively. Fig.~\ref{fig1} displays the errors in parallaxes and PMs with 
respect to $G$ magnitude. In some cases, the catalogue of {\it Gaia} DR2 includes 
parallaxes and PMs with large errors. We did not use the data with errors larger 
than three times the standard deviation from the mean of the errors in parallax and 
PM. The errors in parallaxes and PMs used in this work are, on average, 
about 0.03 mas and 0.05 mas yr$^{-1}$, respectively. Since the 
standard deviations of errors are small (about 0.01 mas for parallax and 0.01 
mas yr$^{-1}$ for PMs), the astrometric data of all stars considered in our study 
are subject to very similar errors. The mean of errors were used as 
a typical error. Stars with negative parallaxes or close companions (duplication flag = 1) 
were also excluded. In addition, stars with parallaxes smaller than five times 
their associated errors were excluded from further analysis. We present the 
full catalogues of our sample stars in Tables~\ref{tab1} and \ref{tab2}, respectively.

In this study, the distance to a given association is determined from the mode 
value of the distance distribution of members. Exclusion of negative or very small 
parallaxes can bias the true distribution of distances particularly for remote 
objects \citep{LHB18}. The distance to our targets can then be underestimated 
by the sample truncation as a mode value of the distance distribution of 
members is expected to be shifted to a smaller distance. To examine its effect 
on the derived distances, the distance distributions of stars were obtained by 
the inversion of the {\it Gaia} parallaxes from the full and truncated samples, 
respectively, and compared with each other. The mode values of these two 
distributions appeared at almost the same distance (less than 10 pc difference). 
This may be because all members of these associations are concentrated in small 
areas relative to their distances. Hence, our sample truncation scheme does not 
significantly influence the distances determined in the present work.

A zero point problem in the {\it Gaia} DR2 parallaxes has been raised. \citet{LHB18} 
estimated a small zero point offset of $-0.03$ mas from the parallax distribution 
of quasars, while \citet{ST18} reported a somewhat larger offset of $-0.08$ mas in 
the zero point from comparison of the parallaxes of known eclipsing binaries with 
those of the {\it Gaia} mission. In addition, \citet{ZPHS19} found a moderate offset 
of $-0.05$ mas from the asteroseismology of red clump stars. These studies indicate 
that the parallaxes measured from the {\it Gaia} mission seem to be slightly underestimated. 
In this paper, we considered the three different zero point offsets as well as the case without 
correction for these offsets. Accordingly, the mean of errors in 
parallaxes increases up to 0.05 mas as a result of the quadratic sum of the 
typical error (0.03 mas) and the errors of the zero point offsets.

In order to probe the kinematic properties of stars in substructures, we 
used the amplitude of a global PM vector defined as below:

\begin{equation}
\mu = \sqrt{\mu^2_{\alpha}\cos^2\delta + \mu^2_{\delta}}
\end{equation}

\noindent The orientation of PMs was also expressed by the 
position angles ($\Phi$) of PM vectors. The errors on these parameters 
were propagated from the errors in PMs. 

\begin{figure*}
   \centering
\includegraphics[width=7.5cm]{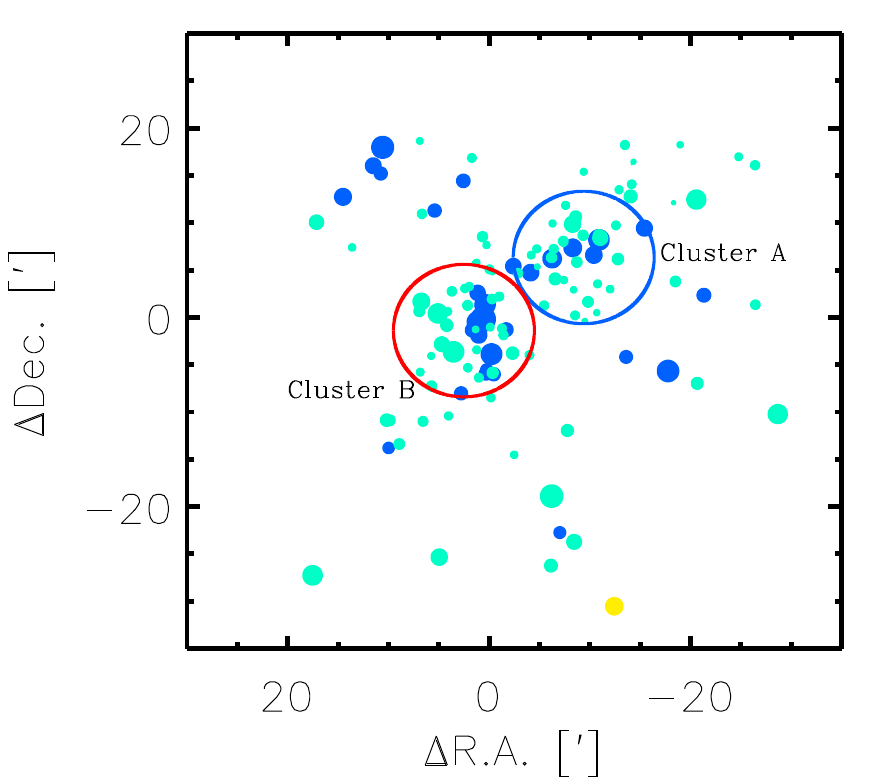} \includegraphics[width=8.4cm]{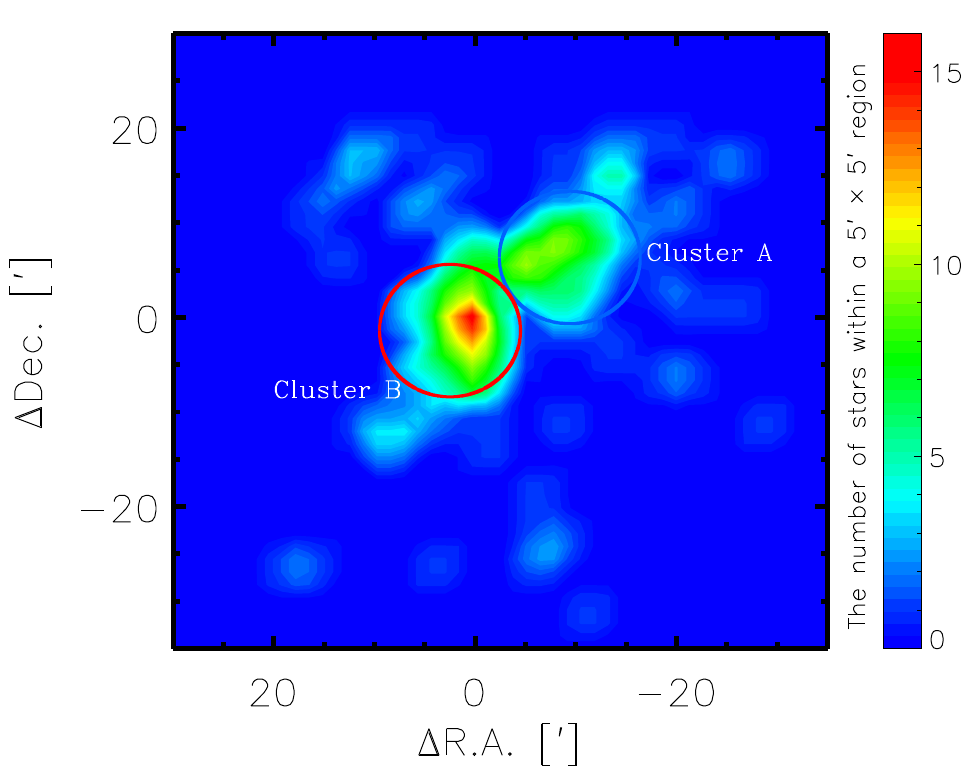}
\includegraphics[width=7.5cm]{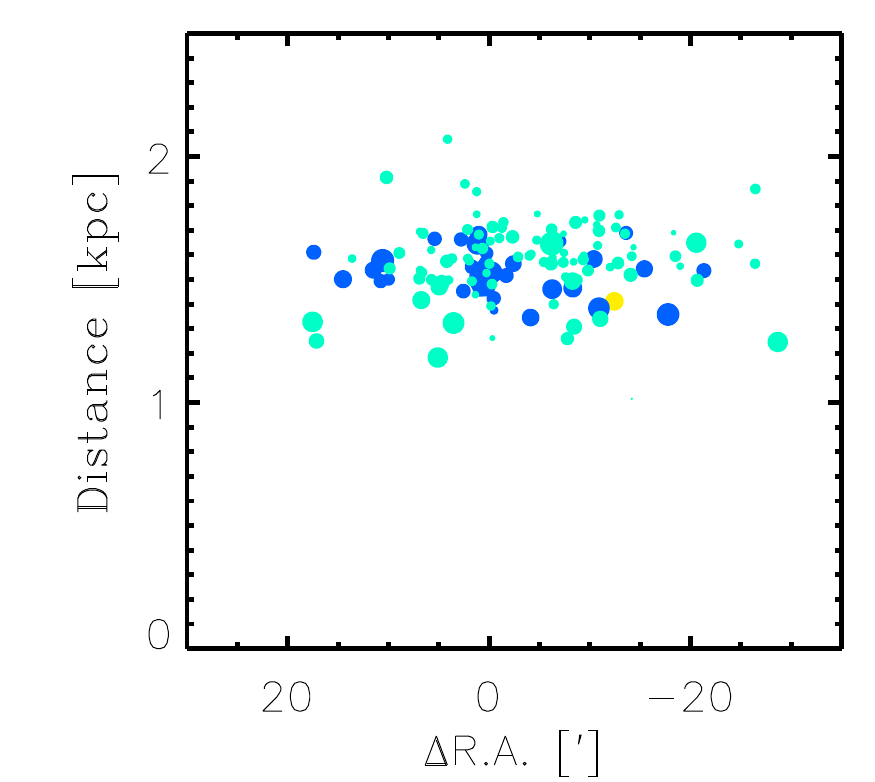}\includegraphics[width=7.5cm]{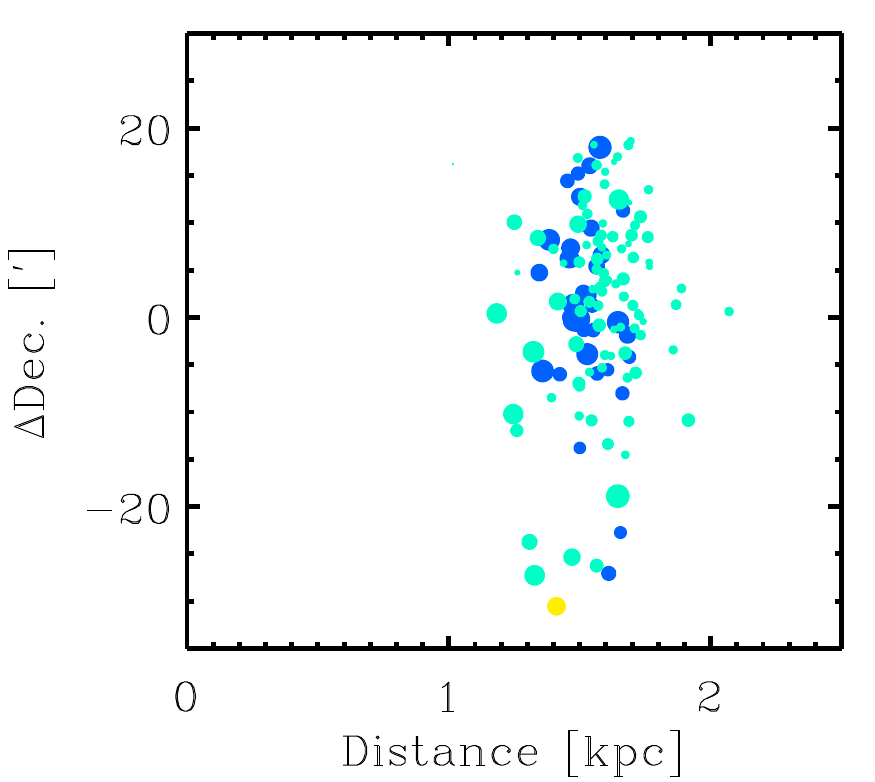}
      \caption{Distribution of stars in Cyg OB2. Upper left-hand panel: Spatial distribution 
of stars. The blue and red circles exhibit the position of Cluster A and Cluster B, respectively. 
Upper right-hand panel: Stellar surface density. The number of stars was counted within areal bins of $5^{\prime} \times 5^{\prime}$. Lower left-hand and lower right-hand panels 
display the distribution of distances to stars with respect to right ascension and declination, 
respectively. The positions of stars are relative to R.A.$ = 20^{\mathrm{h}} \ 
33^{\mathrm{m}} \ 12\fs00$, Dec.$ = 41^{\circ} \ 19^{\prime} \ 1\farcs2$ (J2000). 
Blue, cyan, and yellow dots represent O-, B-type, Wolf-Rayet stars, respectively. The size 
of dots are proportional to the brightness of stars. A moderate offset of $-0.05$ mas 
\citep{ZPHS19} was applied to the parallaxes of stars to obtain their distances. }
         \label{fig3}
 \end{figure*}

\begin{figure*}
   \centering
\includegraphics[width=7.5cm]{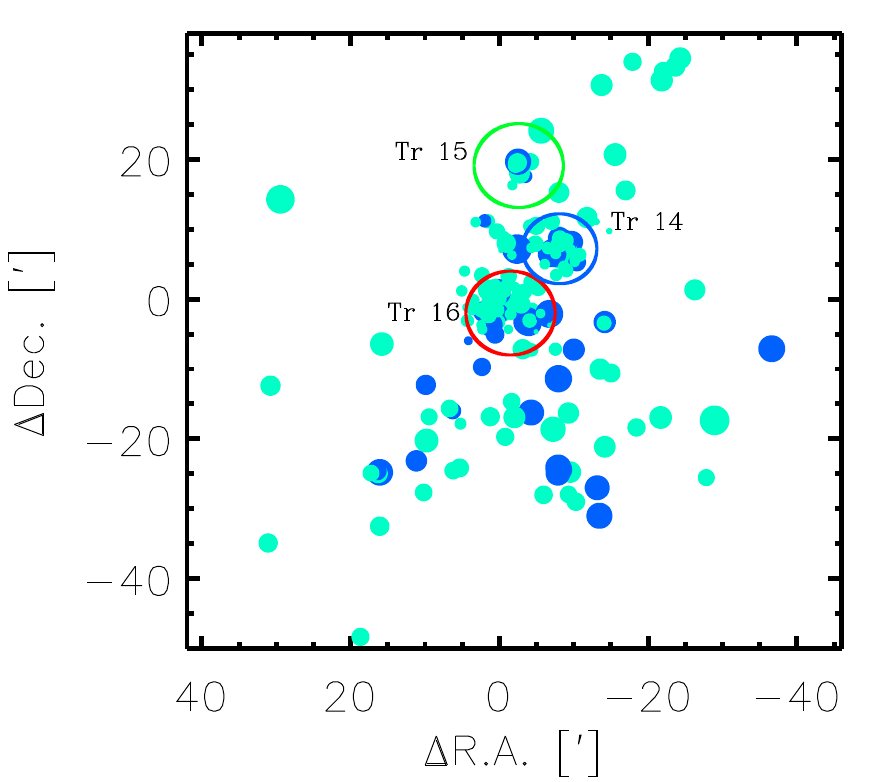} \includegraphics[width=8.4cm]{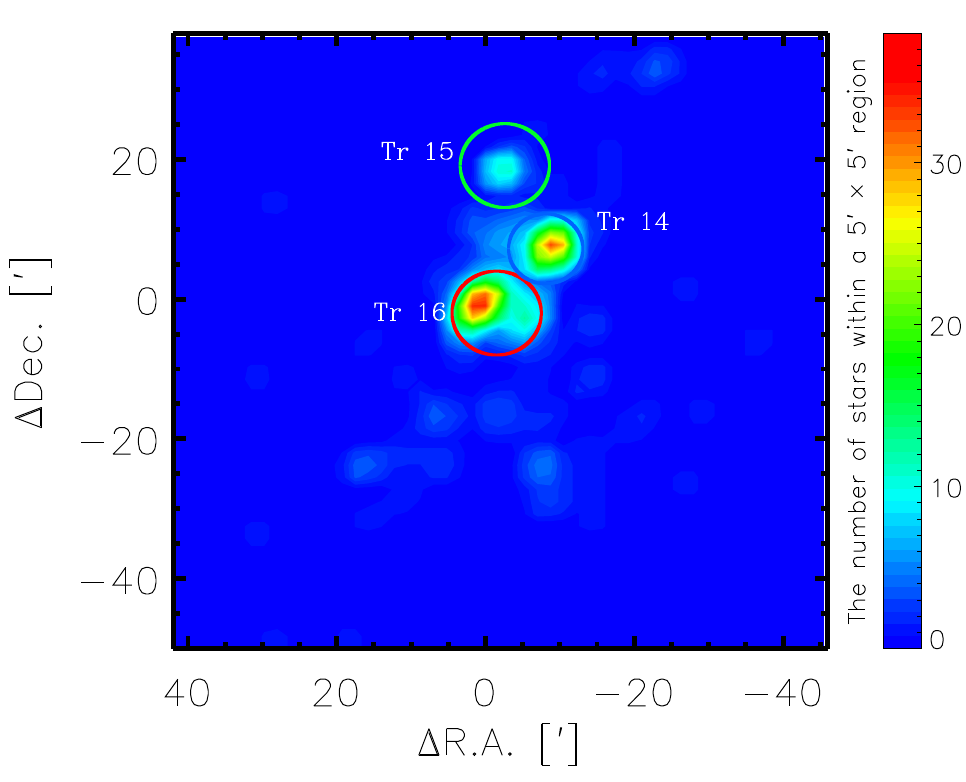}
\includegraphics[width=7.5cm]{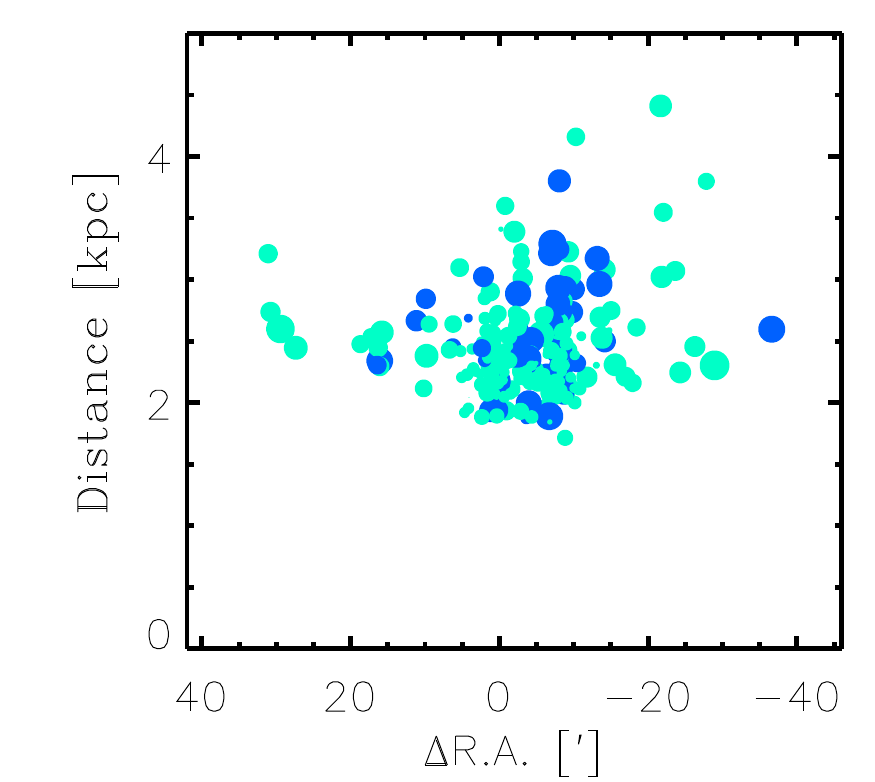}\includegraphics[width=7.5cm]{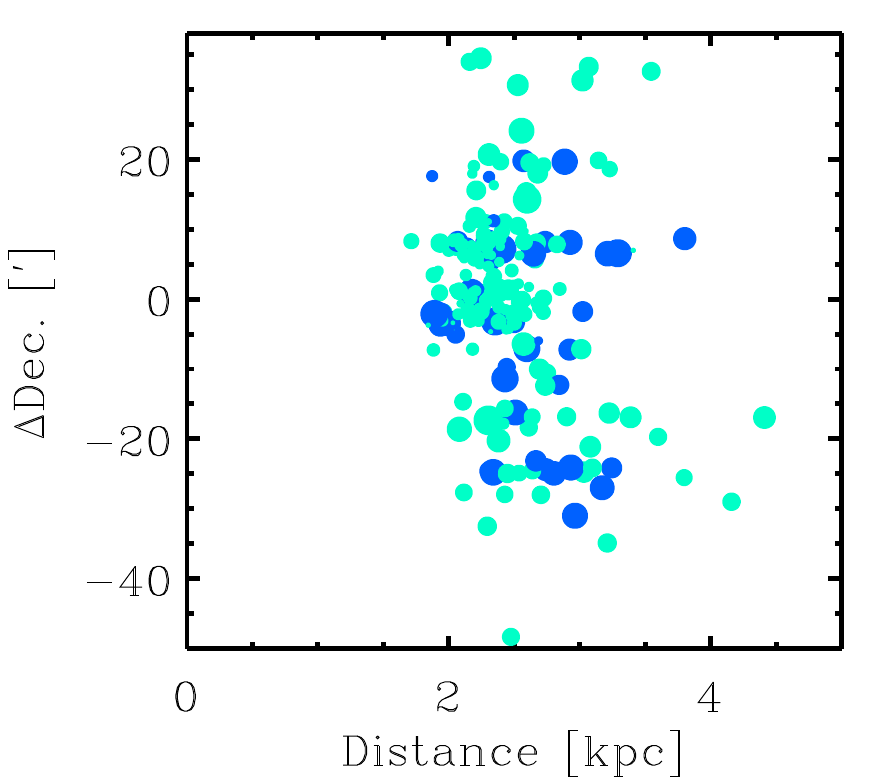}
      \caption{Distribution of stars in Car OB1. The positions of stars are relative 
to $\eta$ Car (R.A.$ = 10^{\mathrm{h}} \ 45^{\mathrm{m}} \ 3\fs55$, 
Dec.$ = -59^{\circ} \ 41^{\prime} \ 3\farcs95$, J2000). The description of each 
panel and the other symbols are the same as in Fig.~\ref{fig3}. The blue, green, 
and red open circles in the upper left-hand panel show the positions of the core 
clusters Tr 14, 15, and 16, respectively. }
  \label{fig4}
 \end{figure*}

\section{Internal structure}
Prior to probing the internal structures of these possible associations, we need 
to check whether or not these are a line-of-sight coincidence 
of several stellar groups. For this purpose, we computed the distances to
 individual stars from the inversion of the {\it Gaia} DR2 parallaxes. Fig.~\ref{fig2} 
exhibits the distance distributions of our sample stars in the direction of 
the two associations. The different systematic offsets were applied to the 
parallaxes in each case. In the case of Cyg OB2, the distributions of distance 
appear to be nearly Gaussian. The distance to Cyg OB2 was estimated to be 
1.6 kpc with a standard deviation of 0.1 kpc. A systematic error of $\pm0.1$ 
kpc can be considered for this result because of the zero point offsets in parallax. 
This result is in reasonable agreement with those derived from previous studies 
\citep{H03,RBS12}. The width of parallax distribution is governed by two terms, 
the intrinsic scatter (radial extent) and the scatter due to measurement errors, i.e., 
$\sigma_{\mathrm{obs}} = \sqrt{\sigma^2_{\mathrm{extent}} + \sigma^2_{\mathrm{error}}}$. 
Typical distance errors converted from the parallax errors (0.03 -- 0.05 mas) are 
comparable to the standard deviation (0.1 kpc). This fact implies that the extent 
of this association along the line-of-sight is smaller than 200 pc.

On the other hand, the distance distributions of stars in the direction of Car OB1 have 
a long tail towards larger distances. A fit by skewed Gaussian profiles yields a peak distance 
of about 2.3 kpc. Systematic errors of $\pm 0.2$ kpc can be considered for the result. 
Car OB1 is located towards the tangent of the Sagittarius-Carina spiral arm, and therefore 
several background OB stars could be observed in the direction of this association. In order 
to check whether or not the stars further away than 3 kpc are background stars, we investigated 
the distribution of {\it Gaia} parallaxes. As a result, it appears to be a single Gaussian 
profile, not a skewed Gaussian profile, which implies that the asymmetric distribution 
at large distance comes from the inversion of the parallax to obtain distance \citep{BRF18}. The results 
obtained from the peaks in the parallax [$0.412\pm0.035$ (sys.) mas which is equivalent 
to $2.4\pm0.2$ (sys.) kpc] and distance distributions are consistent within the systematic 
errors. The Gaussian width is about 0.05 mas, which is equivalent to 0.3 kpc. If we adopt 
a typical error of 0.03 mas, the radial size of Car OB1 is then about 420 pc. However, this 
is an upper limit because the distribution of parallaxes can be entirely governed by the 
uncertainties if the error on the systematic error (0.03 mas) from \citet{ST18} is considered. 

\begin{figure*}
   \centering
\includegraphics[width=15cm]{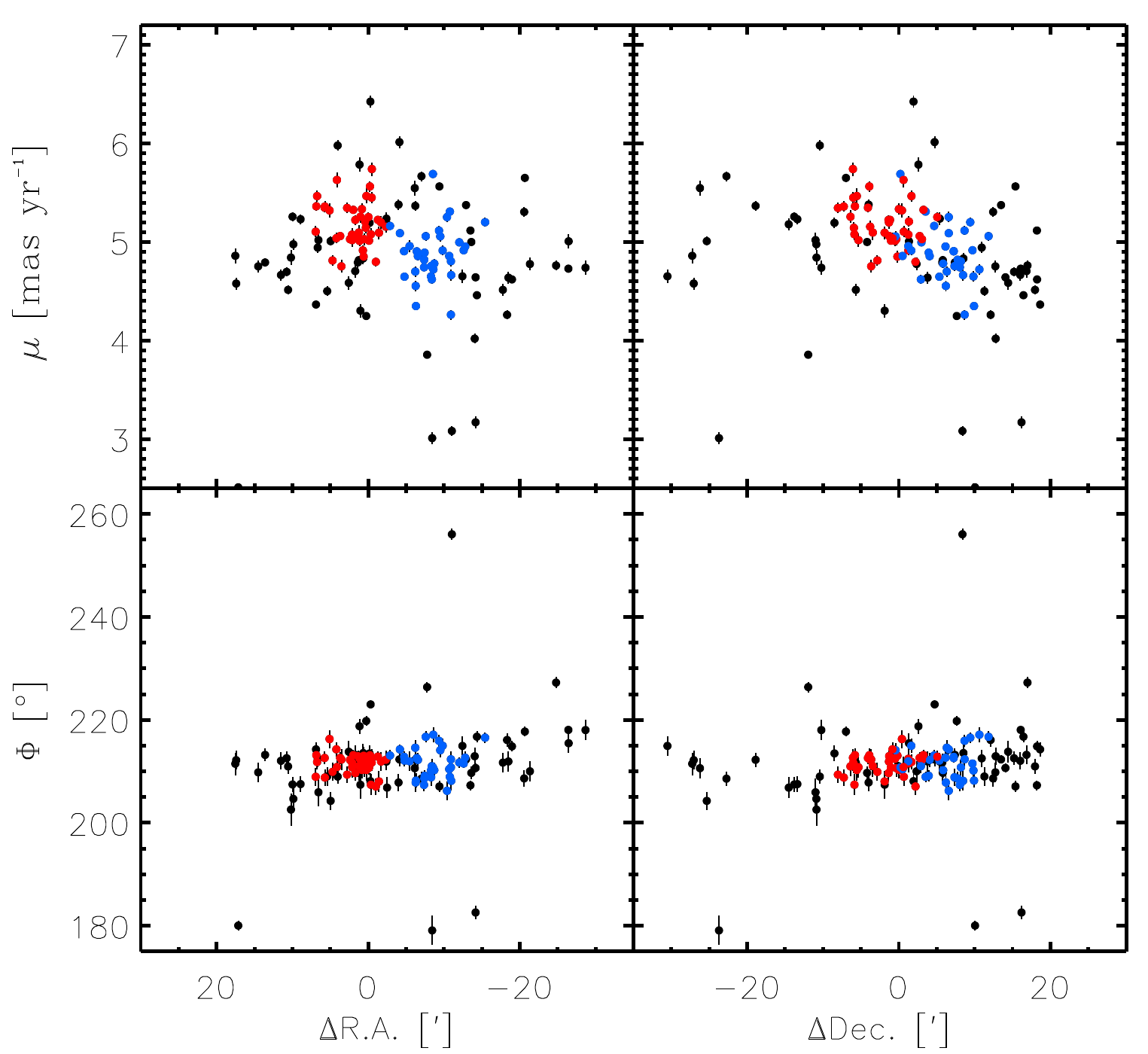} 
      \caption{PM distributions of stars in Cyg OB2. Upper panels display the 
amplitude of PMs ($\mu$) with respect to right ascension and declination, respectively, while 
lower panels exhibit the orientations of PMs ($\Phi$) along each equatorial coordinate. Blue, 
red, and black dots represent Cluster A, Cluster B, and the halo stars, respectively.}
         \label{fig5}
 \end{figure*}

\begin{figure*}
   \centering
\includegraphics[width=15cm]{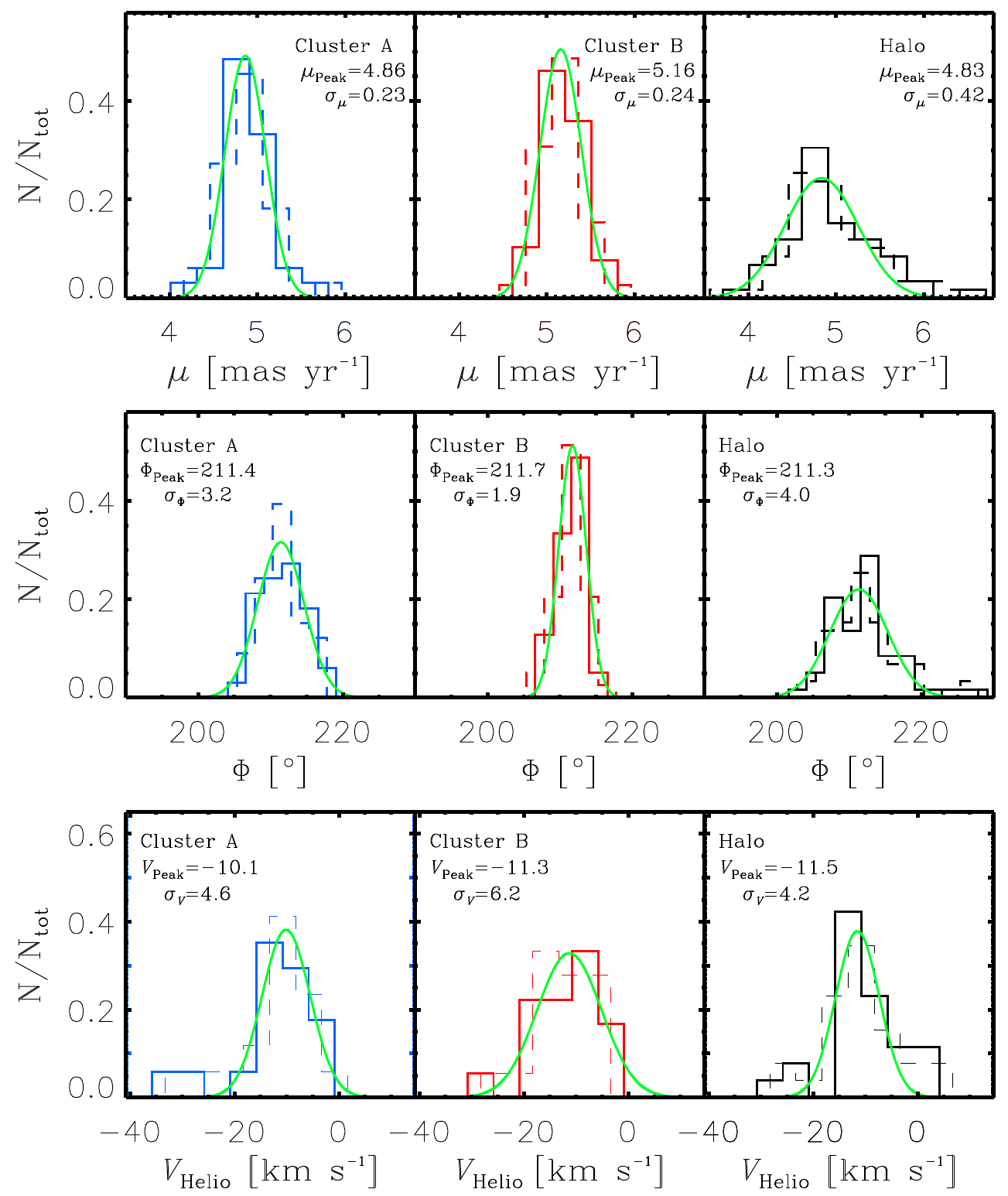} 
      \caption{Distributions of PMs and RVs of stars 
in Cyg OB2. Bin sizes of 0.3 mas yr$^{-1}$, 2.5$^{\circ}$, and 5 km s$^{-1}$ 
were adopted for obtaining each distribution. Two histograms are presented, based on two 
positions for bin centres (solid and dashed lines), to show the impact of the binning. The names 
of stellar groups and the best-fit parameters are labelled at the top of each panel. The curves 
drawn by green solid lines are the best-fit Gaussian distributions.}
         \label{fig6}
 \end{figure*}

Such a large extent can be understood in the context of star formation. Indeed, 
the young open cluster NGC 281 and two H$_2$O maser sources, IRAS 00259+5625 
and IRAS 00420+5530, are part of an expanding ring of molecular clouds, the 
so-called Megeath's ring \citep{MBD02,MBD03}. Their parallaxes measured by using 
Very Long Baseline Interferometry Exploration of Radio Astrometry indicate that the 
three objects are lying along the line-of-sight (see figure 4 in \citealt{SHH08} and 
figure 5 in \citealt{SSM14}). The diameter of the expanding ring derived from the 
distance difference among them is about 0.7--1.9 kpc \citep{SHH08}. This result 
implies that star formation can take place on a scale of several hundred parsecs. 
Our result naturally explains the discrepancy in distance among different studies of Car OB1 
(2.2--2.9 kpc, \citealt{AH93,S06,HSB12}).

There are a few foreground OB stars at $d < 1.0 $ kpc and at $d < 1.7$ kpc in the 
direction of Cyg OB2 and Car OB1, respectively. A probable background star towards 
Cyg OB2 was also found. We list these OB star 
candidates in Table~\ref{tab3} and do not use them in further analysis. Except for 
these stars, most of the other OB stars are located in the same star forming complex, 
and therefore Cyg OB2 and Car OB1 are real stellar systems not chance alignments.

Figs.~\ref{fig3} and ~\ref{fig4} display the spatial distribution of high-mass 
stars in those associations, respectively. The positions of stars are relative to 
R.A.$ = 20^{\mathrm{h}} \ 33^{\mathrm{m}} \ 12\fs00$, Dec.$ = 41^{\circ} \ 19^{\prime} 
\ 1\farcs2$ (J2000) for Cyg OB2 and R.A.$ = 10^{\mathrm{h}} \ 45^{\mathrm{m}} \ 3\fs55$, 
Dec.$ = -59^{\circ} \ 41^{\prime} \ 3\farcs95$, (J2000) for Car OB1. In Cyg OB2, 
there are two groups of stars with a high-stellar density at ($\Delta \alpha$, 
$\Delta \delta$) $\sim$ ($-9\farcm4$, $6\farcm3$) and ($2\farcm5$, $-1\farcm4$). 
A low-density halo extending up to about 20$^{\prime}$ surrounds these clusters. The presence 
of these three substructures is confirmed in the surface stellar density map (upper right-hand 
panel of Fig.~\ref{fig3}), where the number of stars was counted within areal bins 
of $5^{\prime} \times 5^{\prime}$. The western and eastern groups contain up 
to 10 and 16 OB stars per areal bin (1.85 and 2.95 OB stars pc$^{-2}$), respectively, 
while the halo encompasses about $0.3 \pm 0.7$ OB stars per areal 
bin (0.06 OB stars pc$^{-2}$) on average. The peak number density of OB stars 
in these groups are at least 30 times higher than that of the halo. Hereafter, the 
western and eastern clusters are referred to as Cluster A and Cluster B, respectively. 

In Car OB1, several star clusters are distributed over this association 
\citep{FGT11}. Tr 14, 15, and 16 among them appear as the most prominent 
cases, due to their abundant high-mass star content (see also 
upper right-hand panel of Fig.~\ref{fig4}). A low-density halo also surrounds 
the entire association and extends to 30$^{\prime}$ along right ascension 
and declination axes. These clusters contain up to 36, 13, and 38 OB stars 
within the areal bin (3.22, 1.16, and 3.40 OB stars pc$^{-2}$), respectively, 
while the mean stellar density of the halo is about $0.3\pm0.9$ OB stars per 
areal bin (0.03 OB stars pc$^{-2}$). For early-type stars, the peak number 
densities of these clusters are similar to those of the clusters in Cyg OB2. Note 
that the number densities of OB stars obtained in this work may be 
a lower limit because later B-type stars and stars with a duplication 
flag in the {\it Gaia} catalogue were not used in this analysis.

A common property of both associations is that they consist of a few 
high-density core clusters and a low-density halo. This structural feature was 
also seen in other associations, such as Cep OB3, Ori OB1, Lac OB1, 
and Cas OB6 \citep{B64,KAG08,SBC17}. This feature could then be a relic of the 
formation of OB associations, which would imply that there may be a common 
process controlling their formation.

\begin{figure*}
   \centering
\includegraphics[width=15cm]{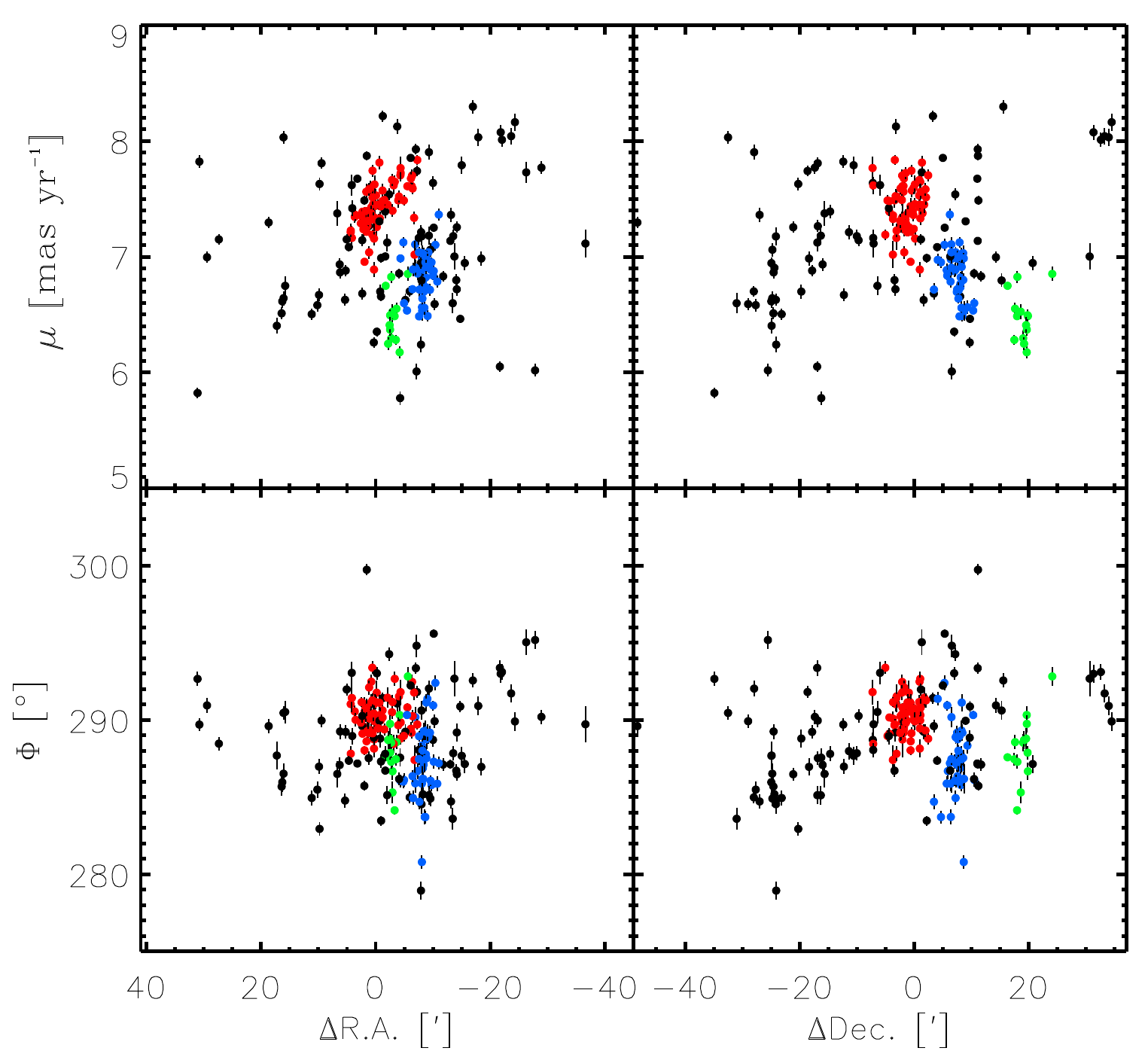} 
      \caption{PM distributions of stars in Car OB1. Blue, green, 
red, and black dots represent Tr14, 15, 16, and the halo stars, respectively.}
         \label{fig7}
 \end{figure*}

\begin{figure*}
   \centering
\includegraphics[width=15cm]{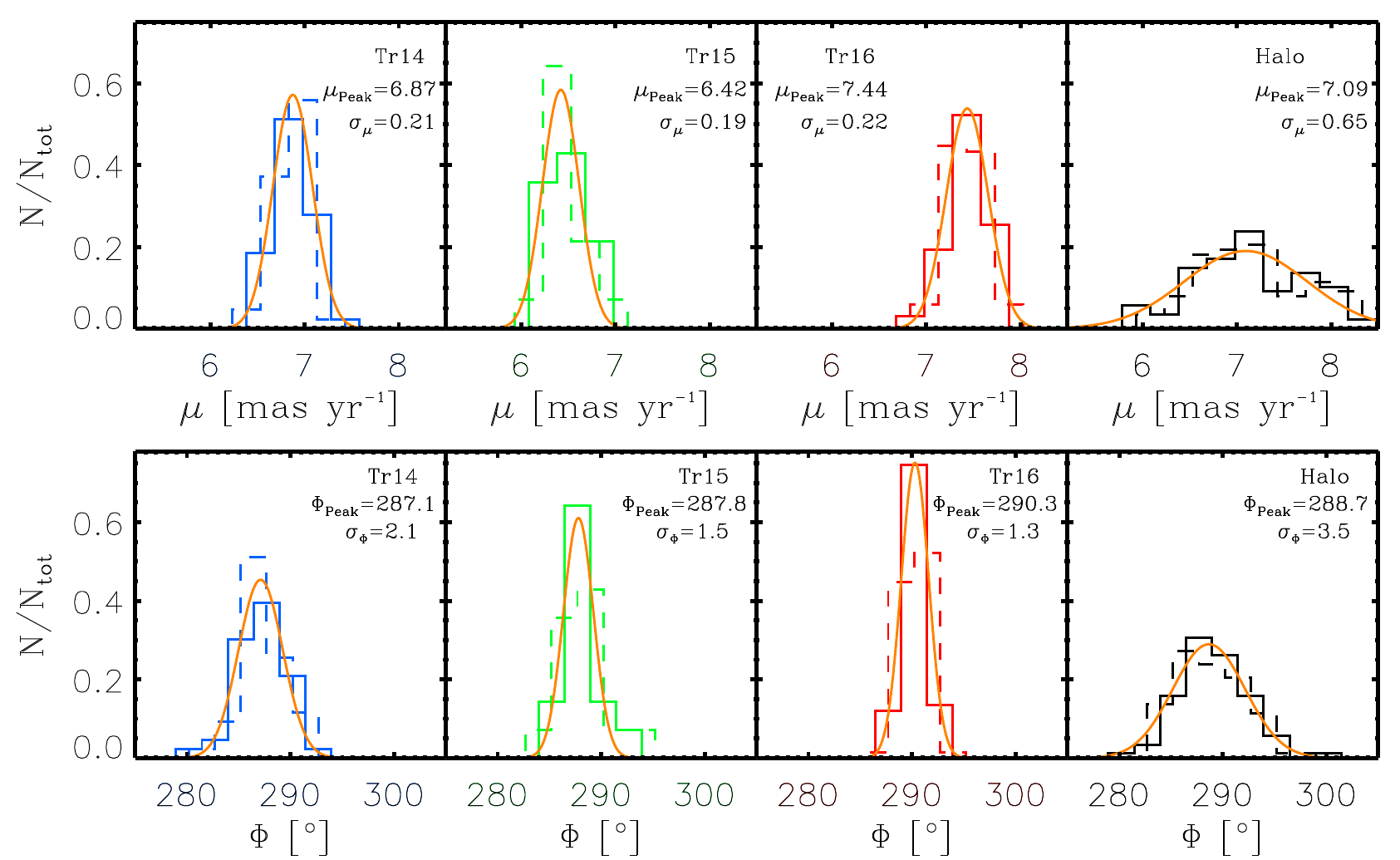} 
      \caption{Distributions of PMs of stars in Car OB1. The bin sizes of 0.3 mas yr$^{-1}$ 
and 2.5$^{\circ}$ were adopted for obtaining each distribution. The names 
of stellar groups and the best-fit parameters are displayed at the top of each panel. The 
curves drawn by orange solid lines are the Gaussian distributions adopting the best-fit 
parameters.}
         \label{fig8}
 \end{figure*}

\begin{figure*}
   \centering
\includegraphics[width=15cm]{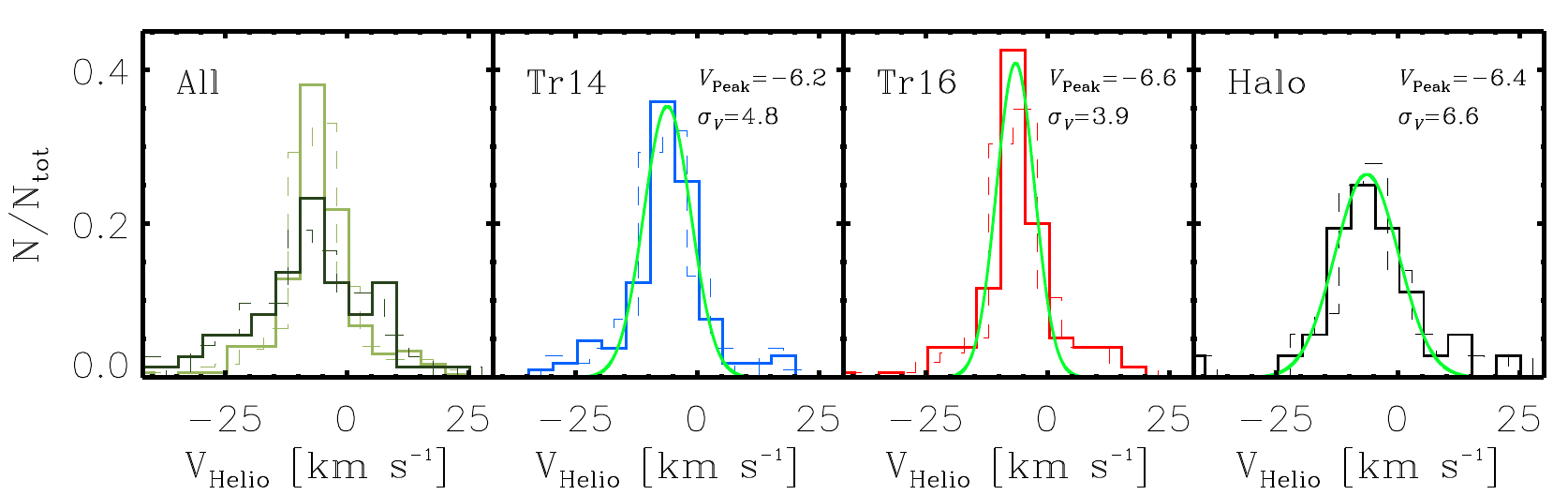} 
      \caption{Distributions of RVs of stars in Car OB1. A bin size of 5 km s$^{-1}$ 
was adopted for obtaining each distribution. In the left-hand panel, dark and 
light histograms display the RV distributions of OB stars and late-type stars from 
\citet{DKJ17} in the entire region, respectively. The other panels exhibit the RV 
distributions of low-mass stars in each cluster (except Tr 15) or the halo. The green 
curves represent the best-fit Gaussian distributions. }
         \label{fig9}
 \end{figure*}

\section{Kinematic substructures}
In order to better characterise the formation process of these associations, we 
probe the kinematics of each substructure. In this section, the criteria of member 
selection are addressed, and the overall properties of individual substructures 
are described using the distributions of $\mu$, $\Phi$, and RVs of members.

\subsection{Cyg OB2}
Fig.~\ref{fig5} displays the distributions of the $\mu$ and $\Phi$ of stars in Cyg 
OB2 with respect to right ascension and declination. Most of the stars are 
moving in the $\mu$ range of 4 mas yr$^{-1}$ to 6 mas yr$^{-1}$. In order to select 
the probable members of Cluster A and Cluster B, we first use the regions assigned 
to these clusters as shown in the upper left-hand panel of Fig.~\ref{fig3}, where 
the centre positions and radii of the regions were visually set to include as many 
probable cluster members as possible from the spatial distribution of stars. Note that 
the positions and radii used in the criteria are thus not the physical centre and size of 
these clusters. The $\mu$ and $\Phi$ of stars in different regions were then plotted 
in Fig.~\ref{fig5} by different colours. Stars with $\mu$ and $\Phi$ values within three 
times the standard deviations from the mean values were selected as members of 
given clusters. The criteria for membership to Cluster A and Cluster B are: 

\begin{enumerate}
\item Cluster A -- R.A. = $20^{\mathrm{h}} 32^{\mathrm{m}} 21\fs89$, \\ 
\indent \indent \indent \indent \indent \indent \indent \indent Dec. = $+41^{\circ} 25^{\prime} 20\farcs8$, J2000, \\
\indent \indent \indent \indent \indent \indent \indent \indent  radius = 7$^{\prime}$\\
\item Cluster B -- R.A. = $20^{\mathrm{h}} 33^{\mathrm{m}} 25\fs31$, \\ 
\indent \indent \indent \indent \indent \indent \indent \indent Dec. = $+41^{\circ} 17^{\prime} 37\farcs2$, J2000, \\
\indent \indent \indent \indent \indent \indent \indent \indent  radius = 7$^{\prime}$\\
\item $\langle \mu \rangle - 3\sigma_{\mu} < \mu < \langle \mu \rangle + 3\sigma_{\mu}$ \\
\item $\langle \Phi \rangle - 3\sigma_{\Phi} < \Phi < \langle \Phi \rangle + 3\sigma_{\Phi}$ \\
\end{enumerate}

\noindent where $\langle \mu \rangle$, $\sigma_{\mu}$, $\langle \Phi \rangle$, and 
$\sigma_{\Phi}$ represent the mean values and standard deviations in 
$\mu$ and $\Phi$ for a given cluster. These values were iteratively 
computed until outliers were completely removed from the subsample. Stars not 
satisfying these criteria were considered as the halo stars. {The numbers of 
members in Cluster A, Cluster B, and the halo are 33, 39, and 59, respectively.}

The centre and size of each substructure were determined using the selected 
members. The centre positions of each substructure quoted in Table~\ref{tab4} 
were obtained from the median coordinates of the selected members. It is 
worth noting that these centre positions well match those used in the criteria for 
member selection. In practice, it is difficult to define the area of the substructures, 
particularly the extent of the halo. Here, the largest distances among members 
in the clusters and halo were adopted as their maximum diameters for simplicity. 
The diameters of Cluster A, Cluster B, and the halo are $6.2 \pm 0.4$, $6.3 \pm 0.4$, 
and $28.7 \pm 1.8$ pc at 1.6 kpc, respectively, where their errors were propagated 
from the systematic errors on distance. 

These clusters have slightly different amplitudes in PM. The members of 
Cluster A have $\mu$ values in the range of 4.2 to 5.8 mas yr$^{-1}$, and 
the $\mu$ values of the Cluster B members are confined between 4.7 and 
5.8 mas yr$^{-1}$. The orientations of their PMs appear to be globally well 
aligned at about 210$^{\circ}$. It means that the star clusters and the halo 
stars are globally moving towards almost the same direction. 

The upper and middle panels of Fig.~\ref{fig6} exhibit the $\mu$ and 
$\Phi$ distributions of stars belonging to Cluster A, Cluster B, and halo, 
respectively. They were fitted by Gaussian profiles, and the best solutions 
(the mean and dispersion) are quoted in each panel. The global PM of Cluster A 
appears smaller than that of Cluster B. The difference is about 0.30 mas 
yr$^{-1}$ (equivalent to 2.3 km s$^{-1}$), which is larger than three times 
the mean of errors (0.05 mas yr$^{-1}$). Stars in the halo are moving, on average, 
at 4.83 mas yr$^{-1}$, which is similar to that of Cluster A but they have 
a dispersion almost twice larger than those found for Cluster A and Cluster B. 
Similarly, the halo is presenting a larger dispersion in $\Phi$. Together with the large 
dispersion in $\mu$, this may indicate that the halo stars suffer from larger 
contributions from random motions than stars in Cluster A and Cluster B. 

We further probed the kinematic substructures of Cyg OB2 along the line 
of sight by using the RVs of 61 out of 102 stars in \citet{KKK07}. 
Spectroscopic binary stars (26) and binary candidates (15) were not used. The lower 
panels of Fig.~\ref{fig6} display the RV distribution of stars in Cluster A, 
Cluster B, and halo, respectively. The mean RVs and dispersions were 
obtained by the Gaussian profile fitting to each RV distribution. The RVs of stars 
in the substructures seem not to be significantly different from each other: 
Cluster A, Cluster B, and the halo have mean RVs of $-10.1$, $-11.3$, and 
$-11.5$ km s$^{-1}$ with different dispersions of 4.6, 6.2, and 4.2 km s$^{-1}$, 
respectively. If Cluster A and Cluster B have isotropic velocity components in 
the three-dimensional space, the velocity difference between these clusters along 
the line-of-sight is about 1.6 km s$^{-1}$ (or $V_{2\mathrm{D}} = 
\sqrt{2} V_{1\mathrm{D}}$ = 2.3 km s$^{-1}$, where $V_{1\mathrm{D}}$ and 
$V_{2\mathrm{D}}$ are a one-dimensional velocity and two-dimensional velocity 
calculated from $\Delta\mu \ \times$ distance, respectively). Because of the large velocity 
dispersions, it would be difficult to strongly argue that this difference 
can be detected in the RV distributions despite the fact that the RV difference between 
these clusters is about 1 km s$^{-1}$. We present the overall properties of the 
substructures in Cyg OB2 in Table~\ref{tab4}.

\begin{figure*}
   \centering
\includegraphics[width=7.5cm]{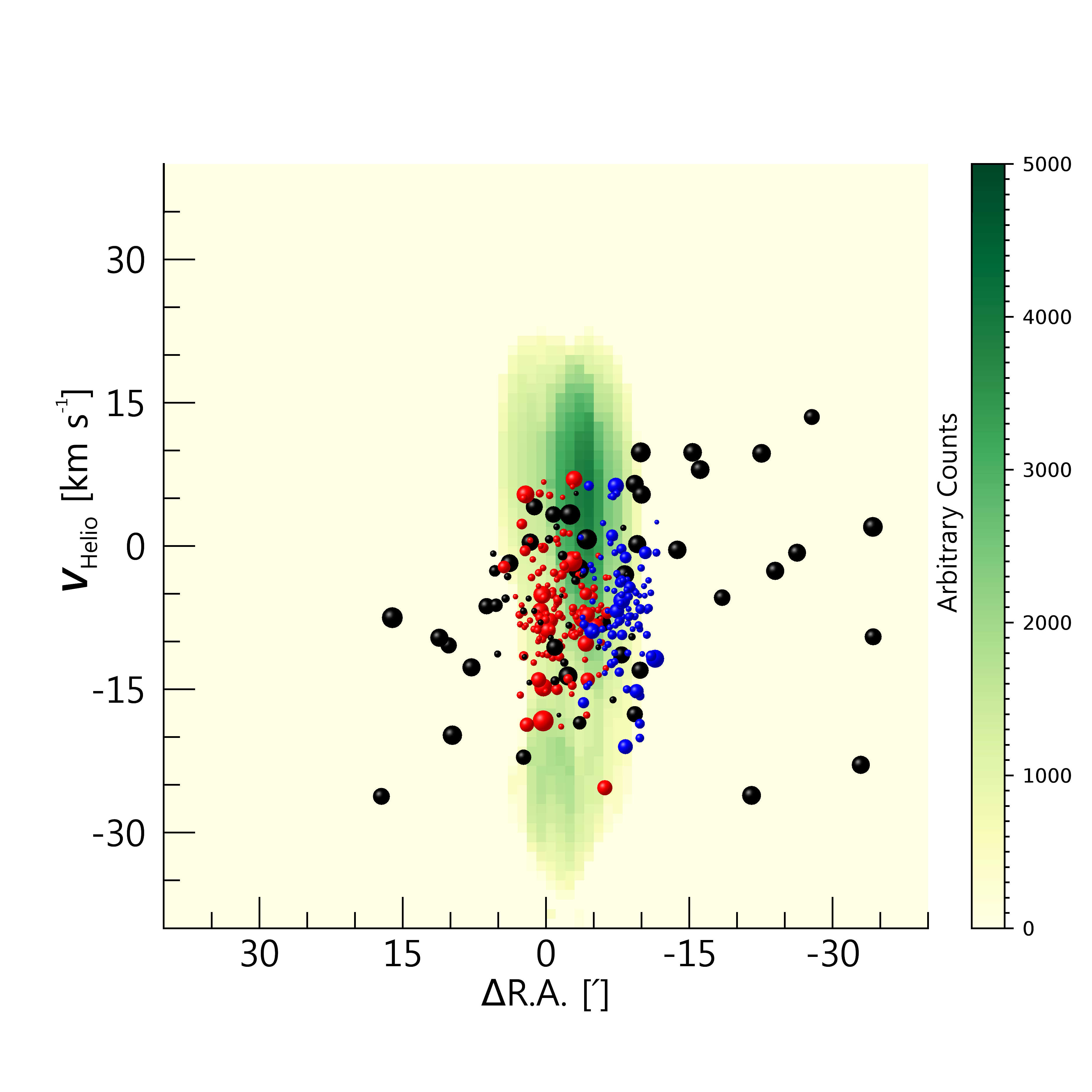}\includegraphics[width=7.5cm]{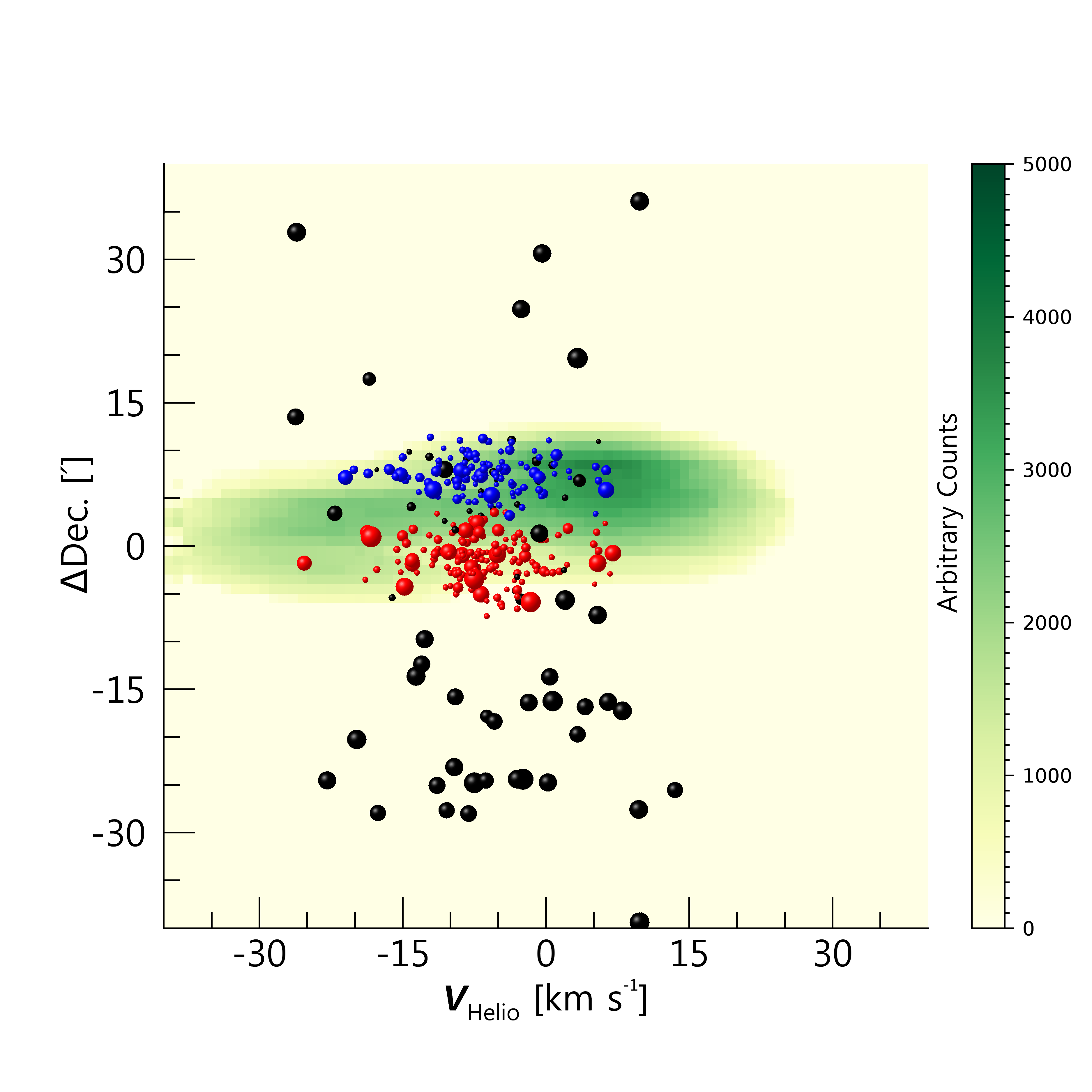}    
      \caption{Position-velocity diagrams of ionised gas and stars along 
right ascension (left) and declination (right). The distribution of ionised gas was 
traced by the forbidden line [N {\scriptsize \textsc{II}}] $\lambda$6584. Red, blue, and 
black spheres denote the members of Tr 14, 16, and the halo, respectively, and their size 
is proportional to the brightness of individual stars in the $G$ band. }
         \label{fig10}
 \end{figure*}

\subsection{Car OB1}
Fig.~\ref{fig7} displays the distributions of the $\mu$ and $\Phi$ of OB stars 
in Car OB1 with respect to their positions. The PMs of these stars range 
from 5.5 mas yr$^{-1}$ to 8.3 mas yr$^{-1}$ in $\mu$, and between 276$^{\circ}$ and 
300$^{\circ}$ in $\Phi$. The kinematics of stars is well correlated with 
the positions of stars. In the same way as for Cyg OB2, we first selected 
the probable members of Tr 14, 15, and 16 within three circular regions as below 
(see the upper left-hand panel of Fig.~\ref{fig4}); 

\begin{enumerate}
\item Tr 14 -- R.A. = $10^{\mathrm{h}} 43^{\mathrm{m}} 59\fs59$, \\ 
\indent \indent \indent \indent \indent \indent  Dec. = $-59^{\circ} 33^{\prime} 51\farcs9$, J2000, \\
\indent \indent \indent \indent \indent \indent   radius = 5$^{\prime}$\\
\item Tr 15 -- R.A. = $10^{\mathrm{h}} 44^{\mathrm{m}} 43\fs14$, \\ 
\indent \indent \indent \indent \indent \indent  Dec. = $-59^{\circ} 21^{\prime} 57\farcs9$, J2000, \\
\indent \indent \indent \indent \indent \indent   radius = 6$^{\prime}$\\
\item Tr 16 -- R.A. = $10^{\mathrm{h}} 44^{\mathrm{m}} 51\fs65$, \\ 
\indent \indent \indent \indent \indent \indent  Dec. = $-59^{\circ} 43^{\prime} 03\farcs9$, J2000, \\
\indent \indent \indent \indent \indent \indent   radius = 6$^{\prime}$\\
\end{enumerate}

\noindent 
The $\mu$ and $\Phi$ of stars are plotted in Fig.~\ref{fig7} by different colours. 
Stars with $\mu$ and $\Phi$ values within three times the standard deviations 
from the cluster mean values were selected as members (see the criteria iii 
and iv used for Cyg OB2). Stars not fulfilling these criteria were assumed to form 
the halo population. The numbers of members in Tr14, 15, 16, and the halo are 
43, 14, 67, and 88, respectively. Note that several OB stars of the other clusters, such as 
Bochum 10, 11, Collinder 232, and 228, may be included in this halo population 
because we only considered the most prominent clusters Tr 14, 15, and 16 in 
this paper. The centre positions of these clusters given in Table~\ref{tab4} were 
also adopted from the median coordinates of members. The maximum diameters 
of Tr 14, 15, 16, and the halo were determined to be $5.1 \pm 0.5$, $5.8 \pm 0.5$, 
$7.9 \pm 0.7$, and $68.9 \pm 6.0$ pc at 2.3 kpc, respectively, from the maximum 
distances among the members. 

Fig.~\ref{fig8} displays the distributions of the $\mu$ and $\Phi$ of stars 
in the three clusters and the halo. The best solutions from Gaussian fitting 
to each distribution are labelled in each panel. Tr 15 has the smallest 
$\mu$ (6.42 mas yr$^{-1}$), while Tr 16 has the largest value 
(7.44 mas yr$^{-1}$). The $\mu$ values of the three clusters seem 
to vary with declination (see the upper right-hand panel of Fig.~\ref{fig7}). 
It is also interesting that these clusters have almost the same dispersion 
(about 0.2 mas yr$^{-1}$) in $\mu$. On the other hand, the halo stars 
have a global PM of 7.09 mas yr$^{-1}$ with a larger dispersion (0.65 mas yr$^{-1}$) 
than those of the clusters. Given the typical error of 0.05 mas yr$^{-1}$, 
this difference in kinematics is a true feature. 

Tr 14 ($287\fdg1$), 15 ($287\fdg8$), and 16 ($290\fdg3$) have 
similar orientations of PMs. Again, the star clusters have 
smaller dispersions ($1\fdg3$ -- $2\fdg1$) than the halo stars ($3\fdg5$), 
implying that the members of the clusters have globally well-aligned motions 
towards given directions. 

In 2013 and 2014, \citet{HMP18} observed a total of 115 OB stars spread over 
Car OB1 and measured their RVs. We took their weighted mean values for the 
same stars, where the inverse of the squared error was used as the weight value. 
Some known binary stars were excluded in this analysis. Only a total of 
19 and 54 OB stars for three clusters and the halo are available in this paper, 
respectively. This number is insufficient to statistically investigate their kinematics, 
particularly for the clusters. For this reason, we used the RVs of 
low-mass stars later than A2 derived by \citet{DKJ17}. Note that the low-mass 
stars in Tr 15 were not observed in their study. The boundaries of Tr 14 and 16 
(see above) were used to distinguish the members of these two clusters. 
The left-hand panel of Fig.~\ref{fig9} compares the RVs of OB stars with those 
of late-type stars. These stars have almost the same mean 
velocities, $-6.9 \pm 11.5$ (s.d.) km s$^{-1}$ for OB stars and $-6.4 \pm 4.5$ 
(s.d.) km s$^{-1}$ for late-type stars. The RV distribution of OB stars appears 
broader than that of late-type stars. This is because the measurement errors 
($\langle \epsilon_{\mathrm{RV}}\rangle =$ 6.5 km s$^{-1}$) for OB stars are 
larger than those ($\langle \epsilon_{\mathrm{RV}}\rangle =$ 4.6 km s$^{-1}$) 
for late-type stars on average.

The other panels of Fig.~\ref{fig9} display the RV distribution of late-type 
stars in Tr 14, 16, and the halo. The mean RVs of these two clusters obtained 
from the best-fit Gaussian profiles are about $-6.2$ and $-6.6$ km s$^{-1}$, 
respectively. Those values are not significantly different given the fact that 
the errors on RVs are larger than the discrepancy: these clusters are thus moving at almost the same 
velocities along the line-of-sight. The halo stars are moving at $-6.4 $ km 
s$^{-1}$, on average, which is also close to the mean RVs of Tr14 and 16. However, their 
velocity dispersion (6.6 km s$^{-1}$) again appears larger than those of the 
clusters (4.8 km s$^{-1}$ for Tr 14 and 3.9 km s$^{-1}$ for Tr 16).

\begin{figure*}
   \centering
\includegraphics[width=15cm]{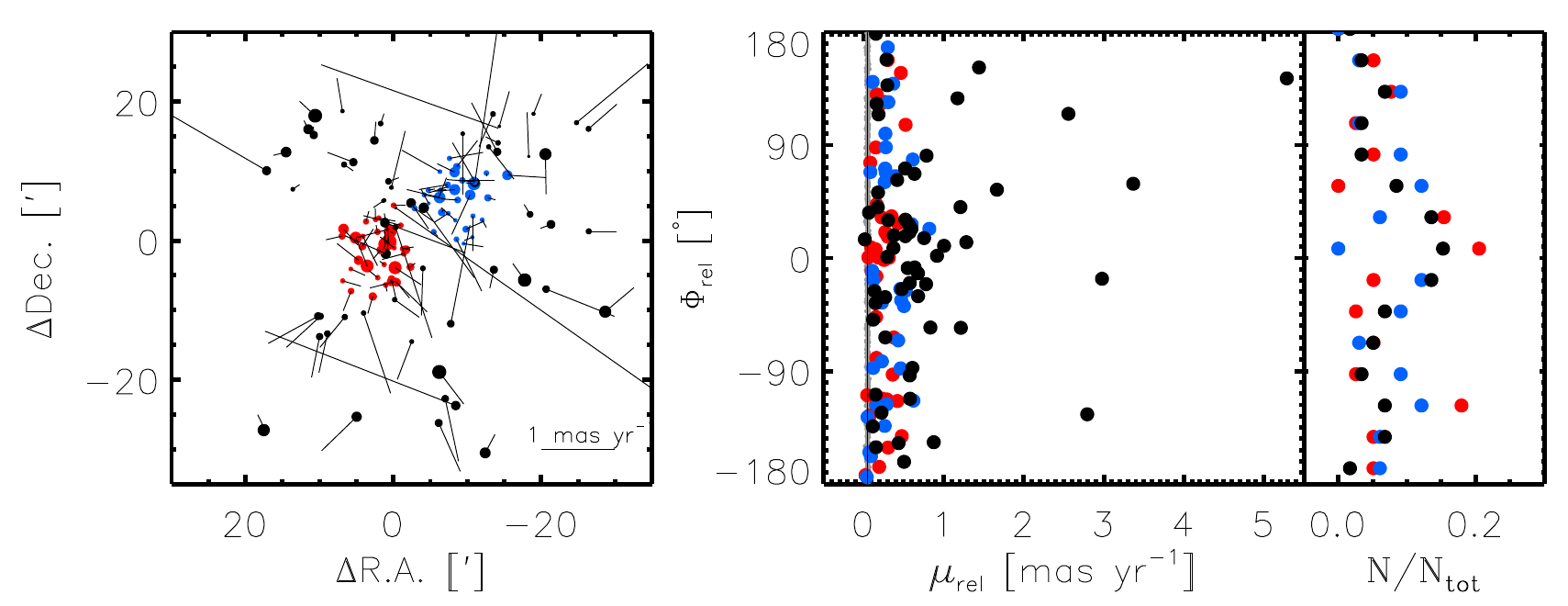} 
\includegraphics[width=15cm]{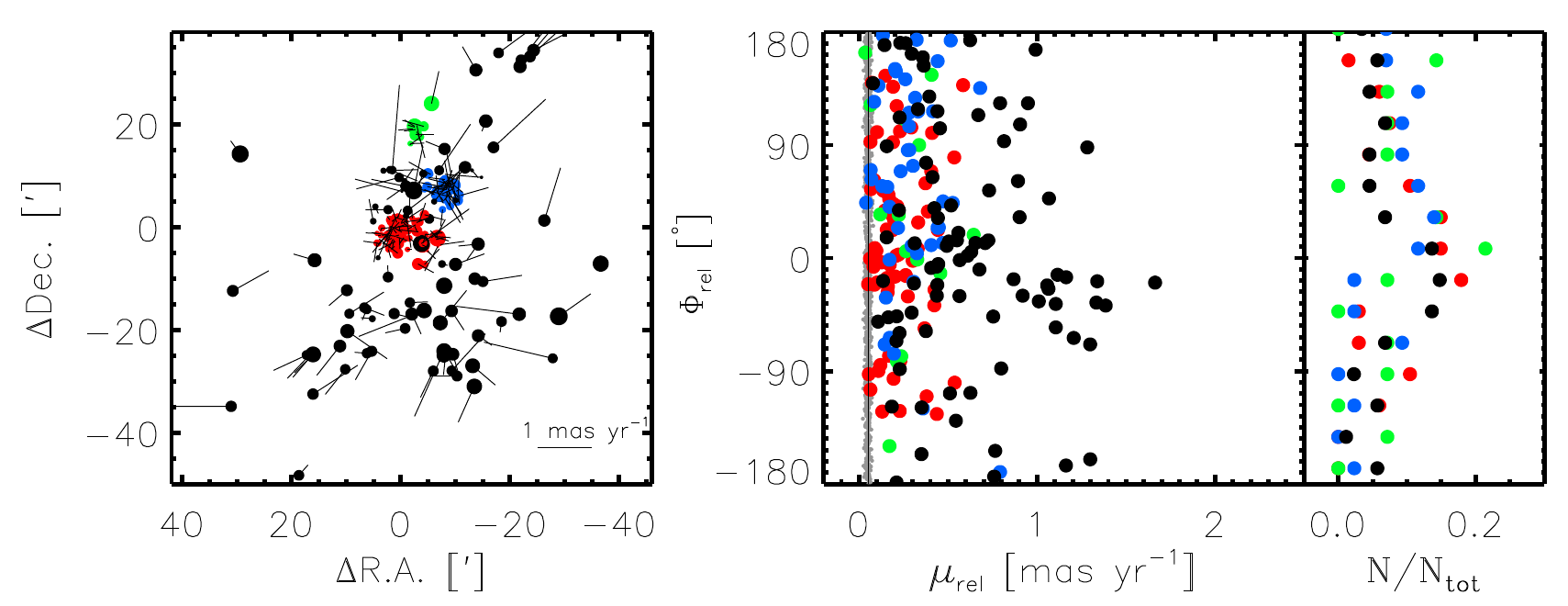}
      \caption{Relative PM vectors of stars in Cyg OB2 (upper) and Car OB1 
(lower). Left panels : spatial distributions of OB stars and their relative PMs. 
The size of dots is proportional to the brightness of individual stars in the 
$G$ band. Straight lines display the PM vectors relative to the median PMs 
of given substructures. Right panels : distribution of $\mu _{\mathrm{rel}}$ 
and $\Phi _{\mathrm{rel}}$. The different colours indicate stars in clusters (red, blue, green) 
or in halos (black). The vertical solid line indicates 
the typical error in $\mu _{\mathrm{rel}}$. Simulated distributions are 
shown by grey dots. The side panels display the distribution of 
$\Phi _{\mathrm{rel}}$, where the histograms were normalised by the total 
number of members belonging to each substructure. }
  \label{fig11}
 \end{figure*}

\begin{figure}
   \centering
\includegraphics[width=7.5cm]{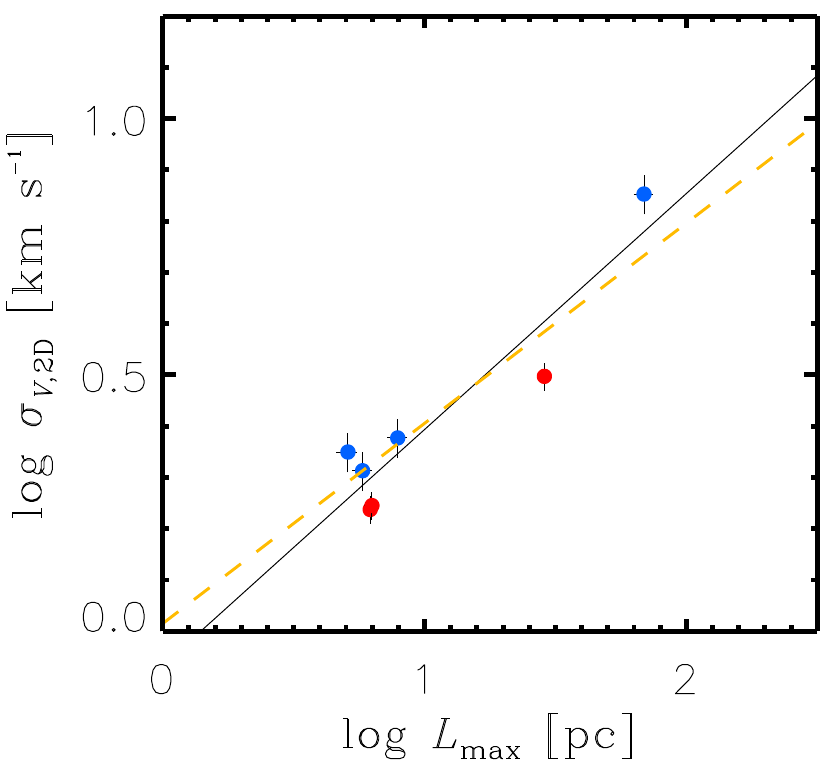}
      \caption{Correlation between the two-dimensional (2D) velocity dispersions and the maximum 
size of substructures. Red and blue dots represent the substructures of Cyg OB2 and Car OB1, 
respectively. The systematic errors in distance were propagated into the 2D velocity 
dispersions and the size of substructures, respectively. The solid line (black) displays the 
result of a linear regression [$\log \sigma_{V,\mathrm{2D}} = 0.46(\pm0.08) \log L_{\mathrm{max}} -0.07 
(\pm 0.09)$], and the dashed line (orange) shows a linear relation [$\log \sigma_{V,\mathrm{2D}} = 0.39 
\log L_{\mathrm{max}} + 0.02$] estimated from the Theil-Sen estimator.}
         \label{fig12}
 \end{figure}

Spectroscopic observations of ionised gas across the Carina nebula were carried 
out by \citet{DBM16}. Their observations allow us to compare the kinematics 
of stars and gas. The optical spectra of the ionised gas show double-peaked 
emission lines of various elements. They provided the best-fit solutions for the 
line profiles, such as RVs, velocity dispersions, and counts. We reconstructed 
a total of 297 synthetic nebular spectra of the forbidden line [N {\scriptsize 
\textsc{II}}] $\lambda$6584 from the best-fit solutions at various nebular 
positions. Subsequently, this information on fibre positions and line profiles 
was interpolated to a regular grid consisting of $80 \times 80 \times 80$ cells 
by applying Delauney triangulation technique as done in \citet{LSB18}. 
Two-dimensional position-velocity diagrams were then obtained by summing the 
counts along right ascension or declination (Fig.~\ref{fig10}), respectively. 
OB and late-type stars with RVs within three times the mean of errors from 
the mean value are also plotted in these diagrams. 
 
The ionised gas of the Carina nebula surrounds Tr 14 and 16 with a shell-like 
appearance in position-velocity space as expected from the double-peaked nature of the 
emission lines \citep{DBM16}. Our sample stars are located between the two shell 
structures. The RV distribution of stars and gas supports our claim that these 
clusters are in the same association. The near-side of the ionised shell is moving 
towards us in the RV range of $-15$ km s$^{-1}$ to $-35$ km s$^{-1}$, while the 
RVs of its far-side range from 0 km s$^{-1}$ to 20 km s$^{-1}$. The median RV 
of each shell was compared with the mean RV of stars. As a result, the 
ionised shell appears to be globally expanding away from the Tr 14 and 16 
clusters at about 17 km s$^{-1}$. The overall properties of the substructures 
in Car OB1 are summarised in Table~\ref{tab4}.

\section{Signature of formation process}
We tested whether or not these associations were formed by expansion of single or multiple 
star clusters. For this purpose, the amplitude of PM vectors ($\mu_{\mathrm{rel}}$) 
relative to the median PMs of given clusters or halos was defined as follows;

\begin{equation}
\mu_{\mathrm{rel}} = \sqrt{(\mu_{\alpha} \cos \delta - \mu_{\alpha ,\mathrm{med}} \cos \delta_{\mathrm{med}})^2+ (\mu_{\delta} - \mu_{\delta ,\mathrm{med}})^2}
\end{equation}

\noindent where the subscript `med' denotes the median PMs of 
given substructures. $\Phi_{\mathrm{rel}}$ was also defined as the angle between 
the radial vector of a star from the centre position of a given 
substructure and its relative PM vector. $\Phi_{\mathrm{rel}}$ 
values of stars should be distributed around 0$^{\circ}$ if the substructures are 
expanding, while the values may appear at either positive or negative $180^{\circ}$ 
in the case of overall contraction. 

Fig.~\ref{fig11} displays the distributions of relative PMs of stars in Cyg OB2 
and Car OB1. In the left-hand panels, PM vectors seem to indicate some expanding 
motions of the halo stars. In the right-hand panels, the distributions are shown 
in more detail. For Cyg OB2, the $\Phi_{\mathrm{rel}}$ of Cluster A (shown in blue) 
appear spread out while those of the halo stars (in black) and of Cluster B (in red) 
display a peak near zero, indicating a contribution from radial expanding motion. 
In Car OB1, results are similar : Tr 14 (in blue) does not show any clear sign of expansion 
whereas Tr 15 (in green), Tr 16 (in red), and the halo (in black) seem to have a true 
low-level expansion. 

Using a Monte-Carlo method, we tested whether or not $\mu_{\mathrm{rel}}$ values 
are dominated by the measurement errors of the {\it Gaia} DR2. The errors in 
$\mu$ are distributed as a Gaussian profile with a standard deviation of 0.01 mas 
yr$^{-1}$ centered at 0.05 mas yr$^{-1}$. A total of 1000 artificial stars 
were generated, and the errors drawn from this error distribution were 
assigned to the $\mu_{\mathrm{rel}}$ of these stars. The $\Phi_{\mathrm{rel}}$ values of 
these stars were randomly generated between $-180^{\circ}$ and 180$^{\circ}$.  
As a result, the mean values of $\mu_{\mathrm{rel}}$ for the substructures in Cyg 
OB2 and Car OB1 are greater than three times the mean value of the simulated 
$\mu_{\mathrm{rel}}$ over all the directions (see grey dots in the right panel 
of Fig.~\ref{fig11}). This simulation therefore seems to indicate the existence of 
several components in kinematics such as low-level expansion, intrinsic random 
motions of stars, and rotation of clusters.

Subgroups of stars in a number of OB associations were found to be 
expanding \citep{MD17,KCS18,KHS19}. Some of the substructures in 
Cyg OB2 and Car OB1 also reveal kinematic properties similar to 
those of the unbound associations. If crossing time of stars is shorter 
than or comparable to their age, the model invoking expansion of clusters 
after rapid gas expulsion \citep{T78,H80,LMD84,KAH01,GB06} could 
explain the formation of Cyg OB2 and Car OB1. The $\mu$ dispersions of star clusters 
are only about 0.2 mas yr$^{-1}$. The halo extends to at least 20$^{\prime}$ from each 
star cluster. The crossing time of stars for this angular distance is about 
6 Myr. This timescale is longer than the age of stars (4--5 Myr for Cyg OB2 
-- \citealt{WDM15}; and 1--3 Myr for Car OB1 -- \citealt{HSB12}). It implies 
that the formation of these two associations cannot be fully explained by the 
cluster expansion. This is consistent with the conclusion of 
\citet{WPGD14,WBD16} although they could not find any signature of 
expansion.

On the other hand, supersonic turbulence plays an important role in the formation 
of hierarchical substructures in gravitationally bound molecular clouds \citep{L81}. 
Recent {\it Herschel} observations showed that nearby molecular clouds form a 
network of filamentary structures \citep{A15}. It seems that turbulent 
flows may be responsible for this structure formation \citep{PJGN01}. Once the 
turbulence is dissipated by shocks, clouds may have different sizes 
and velocity dispersions according to their virial states. This is observed as 
a relation between the velocity dispersions and the size of clumps and molecular 
clouds, i.e. $\sigma \propto L^{\alpha}$, where $\alpha$ ranges from 
0.3 to 0.5 \citep[etc]{L81,S81}. \citet{CBZB05} successfully reproduced structures 
in OB associations by applying this relation to their smoothed particle 
hydrodynamics simulations for star formation in gravitationally unbound clouds. 
If hierarchical star formation driven by turbulence is a more preferable 
explanation than the dynamical evolution of stellar systems, then the observed 
structures and kinematics of OB associations should be similar to those of 
molecular clouds. 

We thus investigated the relation between velocity dispersions and sizes of the 
substructures. The two-dimensional (2D) velocity dispersions ($\sigma_{V,2D}$) of the 
substructures were obtained by $\sigma_{\mu}$ multiplied by distance. 
The errors on $\sigma_{V,2D}$ were propagated from the systematic errors 
in distance. The maximum diameters of each substructure determined in Section 4 
were adopted. Interestingly, Fig.~\ref{fig12} reveals a possible correlation 
between the 2D velocity dispersions and the maximum size of substructures 
in Cyg OB2 and Car OB1. We determined the slope of the correlation using a least-square fitting 
method and found that the velocity dispersions follow a relation $\sigma_{V,2D} \propto 
L_{\mathrm{max}}^{0.46\pm0.08}$. The slope of this correlation was also estimated using a Theil-Sen estimator, leading to a value of 0.39. These two results are consistent 
with each other within the error, and the derived trend is similar to that found in molecular clouds. 
Star formation in turbulent and globally unbound clouds may therefore be a better 
explanation than the expansion of clusters for the formation of Cyg OB2 and Car OB1 (see also the review of 
\citealt{G18}). However, the number of the identified substructures is insufficient 
to reach a definite conclusion in the current state. A systematic survey for more OB 
associations will be able to identify a number of substructures and to confirm this 
relation. Nevertheless, our results are supported by the fact that \citet{KHS19} 
also found a signature of a positive correlation between one-dimensional velocity 
dispersions and half-mass radii of star clusters in different OB associations.

\section{Summary}
We studied the internal structure and kinematics of the two OB associations Cyg OB2 
and Car OB1 in the Galaxy using the high-precision astrometric data from {\it Gaia} DR2 
\citep{gaia18} and RVs from previous studies \citep{KKK07,DKJ17,HMP18}. From the distribution of 
distances, we concluded that these associations are genuine structures rather than 
a line-of-sight coincidence of several stellar groups. They were found to comprise dense core clusters and 
a sparse halo as seen in other OB associations. These substructures reveal different 
kinematic properties from each other in PM. Star clusters with small extent (5 -- 8 parsecs) tend to have 
small dispersions in both amplitude ($\mu$) and orientation ($\Phi$) of PM, while the halo 
stars, spread over a few tens of parsecs, reveal larger dispersions. We also probed the RV 
distributions of stars in each substructure. Stars in Cyg OB2 do not show significant differences 
between clusters and halo in RV. In Car OB1, the velocity dispersion of the 
halo stars appears larger than those of stars within clusters, as for their PM 
distributions.

The relative PMs of stars in star clusters and halo showed that some of these 
substructures reveal a weak signature of expansion. However, the expansion 
of clusters cannot fully explain the formation of these associations given the large 
crossing time of stars, which is longer than their ages. Instead, a correlation between 
the sizes and velocity dispersions of the substructures was 
found, which is reminiscent of the ``size-line width'' relation of molecular clouds 
\citep{L81}. If this correlation was inherited from that of natal clouds, these 
associations might have formed in turbulent giant molecular clouds. However, 
because of small number statistics, this correlation should be confirmed by 
identifying more substructures in other associations and enlarging the sample 
size from the forthcoming {\it Gaia} data.

\section*{Acknowledgements}
The authors thank the anonymous referee for many constructive comments. 
This work has made use of data from the European Space Agency (ESA) mission
{\it Gaia} (https://www.cosmos.esa.int/gaia), processed by the {\it Gaia}
Data Processing and Analysis Consortium (DPAC,\url{https://www.cosmos.esa.int/web/gaia/dpac/consortium}). 
Funding for the DPAC has been provided by national institutions, in particular the 
institutions participating in the {\it Gaia} Multilateral Agreement. This work was 
supported by the National Research Foundation of Korea (NRF) grant 
funded by the Korea government (MSIT) (Grant No: NRF-2019R1C1C1005224), Basic 
Science Research Program through the NRF funded by the Ministry of Education 
of Korea (Grant No. NRF-2017R1A6A3A03006413), and BK 21 plus program through the 
NRF funded by the Ministry of Education of Korea. YN, EG, and GR also acknowledge support 
by the FNRS and by Belspo through the PRODEX contracts.

%%%%%%%%%%%%%%%%%%%%%%%%%%%%%%%%%%%%%%%%%%%%%%%%%%



%\bibliographystyle{mnras}


\begin{landscape}
\begin{table} {\tiny
\caption{Catalogue of high-mass stars in Cygnus OB2.}
\label{tab1}
  \begin{tabular}{cccccccccccccccccc}
  \hline
  \hline
Sq. & $\alpha_{J2000}$  & $\delta_{J2000}$ & Spectral Type & $p$ & $\epsilon (p)$ & $\mu_{\alpha}\cos \delta$  & $\epsilon(\mu_{\alpha}\cos \delta)$ & $\mu_{\delta}$ & 
$\epsilon (\mu_\delta)$ & Dup$^{2}$ & $G$ & $\epsilon (G)$ & $B_p$ & $\epsilon (B_p)$ & $R_p$ & $\epsilon (R_p)$ & $B_p - R_p$ \\
  & (h m s)  & ($^{\circ}$ $^{\prime}$ $^{\prime\prime}$) &  & (mas) & (mas) & (mas yr$^{-1}$)  & (mas yr$^{-1}$) & (mas yr$^{-1}$) & 
mas yr$^{-1}$ &  & (mag) & (mag) & (mag) & (mag) & (mag) & (mag) & (mag) \\
  \hline
   1 & 20:32:03.10 & +41:15:19.9 & WC4                  & 0.5355  & 0.0510   & -2.619  & 0.085    & -4.157  & 0.084 & 0 & 12.8727 & 0.0004 & 14.3121 & 0.0046 & 11.6628 & 0.0015 &  2.6493 \\
   2 & 20:32:06.29 & +40:48:29.7 & WN7o/CE+O7V((f))     & 0.6555  & 0.0415   & -2.664  & 0.066    & -3.813  & 0.069 & 0 & 10.6348 & 0.0010 & 12.0984 & 0.0027 &  9.4580 & 0.0023 &  2.6404 \\
   3 & 20:35:47.10 & +41:22:44.7 & WC6+O8III            & 0.8345  & 0.3676   & -2.916  & 0.628    & -3.875  & 0.786 & 0 & 11.0537 & 0.0008 & 13.0142 & 0.0056 &  9.6917 & 0.0033 &  3.3225 \\
   4 & 20:33:08.78 & +41:13:18.1 & O3If+O6V             & 0.6231  & 0.0667   & -2.685  & 0.095    & -4.601  & 0.107 & 0 & 10.8262 & 0.0012 & 12.1324 & 0.0349 &  9.4787 & 0.0274 &  2.6538 \\
   5 & 20:33:14.16 & +41:20:21.5 & O3If                 & 0.6248  & 0.0298   & -2.724  & 0.051    & -4.438  & 0.052 & 0 &  9.7405 & 0.0005 & 10.7503 & 0.0013 &  8.7606 & 0.0017 &  1.9897 \\
   6 & 20:33:18.02 & +41:18:31.0 & O5III                & 0.5544  & 0.0306   & -2.687  & 0.046    & -4.229  & 0.053 & 0 &  9.4794 & 0.0006 & 10.3727 & 0.0015 &  8.5614 & 0.0013 &  1.8112 \\
   7 & 20:33:10.74 & +41:15:08.0 & O5I+O3.5III          & 0.6012  & 0.0326   & -3.041  & 0.045    & -4.659  & 0.057 & 0 &  9.6060 & 0.0010 & 11.0193 & 0.0015 &  8.4582 & 0.0012 &  2.5611 \\
   8 & 20:34:08.55 & +41:36:59.3 & O5If+B0V             & 0.5810  & 0.0268   & -2.322  & 0.041    & -3.871  & 0.045 & 0 &  9.2414 & 0.0005 & 10.2712 & 0.0015 &  8.2541 & 0.0019 &  2.0172 \\
   9 & 20:33:23.46 & +41:09:12.9 & O5.5V                & 0.0000  & 0.0000   &  0.000  & 0.000    &  0.000  & 0.000 & 0 & 10.8943 & 0.0234 & 12.0022 & 0.0063 &  9.1825 & 0.0060 &  2.8197 \\
  10 & 20:33:13.25 & +41:13:28.6 & O6V                  & 0.5694  & 0.0445   & -2.754  & 0.072    & -4.722  & 0.078 & 0 & 12.2473 & 0.0004 & 13.7639 & 0.0039 & 11.0366 & 0.0020 &  2.7273 \\
\hline
\end{tabular}
\begin{tabular}{@{}l@{}}
Col. (1) : Sequential number. Cols. (2) and (3) : The equatorial coordinates of stars. Col. (4) : Spectral types of stars listed in \citet{WDM15}. Cols. (5) and (6) : Absolute parallax and its standard error. Cols. (7) and (8) : \\
Proper motion in the direction of right ascension and its standard error. Cols. (9) and (10) : Proper motion in the direction of declination and its standard error. Col. (11) : Duplication flag. Cols. (12) and (13) : $G$ magnitude and \\ 
its standard error. Cols. (14) and (15) : $B_p$ magnitude and its standard error. Cols. (16) and (17) : $R_p$ magnitude and its standard error. Col. (18) : $B_p - R_p$ colour index. All the data listed from Col. (5) to Col. (18) were taken \\
from Gaia DR2 \citep{gaia18}.\\

The full table is available electronically.
\end{tabular}}
\end{table}
\end{landscape}

\begin{landscape}
\begin{table} {\tiny
\caption{Catalogue of high-mass stars in Carina OB1.}
\label{tab2}
  \begin{tabular}{cccccccccccccccccc}
  \hline
  \hline
Sq. & $\alpha_{J2000}$  & $\delta_{J2000}$ & Spectral Type & $p$ & $\epsilon (p)$ & $\mu_{\alpha}\cos \delta$  & $\epsilon (\mu_{\alpha}\cos \delta)$ & $\mu_{\delta}$ &
$\epsilon (\mu_\delta)$ & Dup & $G$ & $\epsilon (G)$ & $B_p$ & $\epsilon (B_p)$ & $R_p$ & $\epsilon (R_p)$ & $B_p - R_p$ \\
  & (h m s)  & ($^{\circ}$ $^{\prime}$ $^{\prime\prime}$) &  & (mas) & (mas) & (mas yr$^{-1}$)  & (mas yr$^{-1}$) & (mas yr$^{-1}$) &
mas yr$^{-1}$ &  & (mag) & (mag) & (mag) & (mag) & (mag) & (mag) & (mag) \\
\hline
   1 & 10:40:12.38 & -59:48:09.7 &  O8V$^{1}$            & 0.3322  & 0.0599   & -6.697  & 0.121     & 2.400  & 0.128 & 0 &  8.0825 & 0.0032 &  8.1060 & 0.0100 &  8.0581 & 0.0094 &  0.0479 \\
   2 & 10:40:30.07 & -59:56:51.1 &  B2III$^{1}$          & 0.3042  & 0.0353   & -6.922  & 0.059     & 3.105  & 0.064 & 1 & 10.5072 & 0.0004 & 10.5697 & 0.0012 & 10.3617 & 0.0010 &  0.2080 \\
   3 & 10:40:31.65 & -59:46:43.6 &  B1.5III$^{1}$        & 0.3784  & 0.0520   & -6.909  & 0.093     & 2.961  & 0.087 & 1 &  8.7903 & 0.0004 &  8.7887 & 0.0016 &  8.7908 & 0.0015 & -0.0021 \\
   4 & 10:40:39.22 & -60:05:35.9 &  B2III$^{1}$          & 0.2773  & 0.0470   & -7.006  & 0.078     & 3.210  & 0.081 & 1 &  9.7862 & 0.0005 &  9.7513 & 0.0012 &  9.8430 & 0.0012 & -0.0917 \\
   5 & 10:41:12.29 & -59:58:24.7 &  B1.5II:$^{1}$        & 0.3814  & 0.0327   & -7.292  & 0.057     & 2.682  & 0.057 & 0 &  7.2207 & 0.0007 &  7.2199 & 0.0022 &  7.2386 & 0.0029 & -0.0187 \\
   6 & 10:41:15.28 & -59:57:45.2 &  B2Ib$^{1}$           & 0.0698  & 0.0283   & -5.694  & 0.051     & 2.666  & 0.047 & 0 & 10.4717 & 0.0004 & 10.7193 & 0.0010 & 10.0700 & 0.0009 &  0.6493 \\
   7 & 10:41:20.24 & -60:06:36.1 &  B2V$^{1}$            & 0.2103  & 0.0315   & -5.447  & 0.056     & 2.560  & 0.052 & 0 & 11.1166 & 0.0010 & 11.1808 & 0.0046 & 10.9472 & 0.0027 &  0.2336 \\
   8 & 10:41:35.44 & -59:39:44.8 &  B1.5V$^{1}$          & 0.3543  & 0.0528   & -7.003  & 0.092     & 3.270  & 0.094 & 0 &  9.8834 & 0.0004 &  9.9028 & 0.0018 &  9.8378 & 0.0017 &  0.0650 \\
   9 & 10:41:54.15 & -59:06:36.3 &  B1V$^{1}$            & 0.3925  & 0.0433   & -7.675  & 0.076     & 2.778  & 0.077 & 0 &  9.6451 & 0.0005 &  9.6505 & 0.0012 &  9.6307 & 0.0012 &  0.0197 \\
  10 & 10:41:55.79 & -59:16:16.4 &  B3III$^{1}$          & 0.1678  & 0.0330   & -7.231  & 0.064     & 3.316  & 0.069 & 1 &  9.7148 & 0.0007 &  9.7866 & 0.0022 &  9.5700 & 0.0019 &  0.2166 \\
\hline
\end{tabular}
\begin{tabular}{@{}l@{}}
Col. (1) : Sequential number. Cols. (2) and (3) : The equatorial coordinates of stars. Col. (4) : Spectral types of stars obtained from previous studies, 1 -- \citet{NBO11}, 2 -- \citet{HSB12}, 3 -- \citet{DKJ17}, \\ 
4 -- \citet{SMM14}, 5 -- \citet{AHPM16}, 6 -- \citet{HMP18}. Cols. (5) and (6) : Absolute parallax and its standard error Cols. (7) and (8) : Proper motion in the direction of right ascension and its standard \\ 
error. Cols. (9) and (10) Proper motion in the direction of declination and its standard error. Col. (11) : Duplication flag. Cols. (12) and (13) : $G$ magnitude and its standard error. Cols. (14) and (15) : $B_p$ magnitude and \\
its standard error. Cols. (16) and (17) : $R_p$ magnitude and its standard error. Col. (18) : $B_p - R_p$ colour index. All the data listed from Col. (5) to Col. (18) were taken from Gaia DR2 \citep{gaia18}.\\

The full table is available electronically.
\end{tabular}}
\end{table}
\end{landscape}

\begin{landscape}
\begin{table} {\tiny
\caption{List of foreground and background star candidates}
\label{tab3}
  \begin{tabular}{lccccccccccc}
  \hline
  \hline
Object & R.A. (2000) & Dec. (2000) & Spectral type & p & $\epsilon$(p) & $\mu_{\alpha}\cos \delta$ & $\epsilon(\mu_{\alpha}\cos \delta)$ & $\mu_{\delta}$ & $\epsilon(\mu_{\delta})$ & $G$ & $\epsilon(G)$  \\
    & (h m s) & ($^{\circ} \ ^{\prime} \ ^{\prime \prime}$) & & (mas) & (mas) & (mas yr$^{-1}$) & (mas yr$^{-1}$)  & (mas yr$^{-1}$) & (mas yr$^{-1}$) & (mag) & (mag)  \\
  \hline
Cyg OB2 &&&&&&&&&&& \\
BD+40 4213   & 20:31:46.00 & +41:17:27.1   & O9.5I$^1$ &  7.1921 & 0.0292 &  -32.407 & 0.050 &  -16.439 &  0.043 & 9.0181 & 0.0003  \\
ALS 15161  & 20:33:10.34 & +41:13:06.4 &   B0V$^1$  &   1.0399 & 0.0459 &   5.728 &  0.062 & -3.398 &  0.087 & 15.8433 & 0.0011 \\
ALS 15175  & 20:33:14.34 & +41:19:33.1 & B5V$^1$ & 0.1867 & 0.0363 & -3.163 &  0.063 & -5.056 & 0.066 & 13.5165 & 0.0003 \\ 
\hline
Car OB1 &&&&&&&&&&& \\
HD 303296  &  10:42:25.04 & -59:09:24.5 &  B1Ve$^2$ & 0.6908 &  0.0690 &  -7.508 & 0.132 & 3.053 & 0.141 & 9.5535  & 0.0007  \\
 HD 93695  & 10:47:44.32 & -59:52:30.9 & B3V$^2$  & 2.3023 & 0.0630 & -14.788 & 0.093 & 0.685 & 0.082 & 6.4318 & 0.0004  \\
\hline
\end{tabular}
\\
\begin{tabular}{@{}l@{}}
Col. (1) : Names of objects. Cols. (2) and (3) : The equatorial coordinates of stars. Col. (4) : Spectral types of stars obtained from previous studies, 1 -- \citet{WDM15} \\ 
and 2 -- \citet{NBO11}. Cols. (5) and (6) : Absolute parallax and its standard error. Cols. (7) and (8) : Proper motion in the direction of right ascension and its standard \\ 
error. Cols. (9) and (10) : Proper motion in the direction of declination and its standard error. Cols. (11) and (12) : $G$ magnitude and its standard error. All the data listed \\ 
from Col. (5) to Col. (12) were taken from Gaia DR2 \citep{gaia18}.\\
\end{tabular}}
\end{table}
\end{landscape}

\begin{landscape}
\begin{table} 
\caption{Properties of the substructures in Cygnus OB2 and Carina OB1.}
\label{tab4}
  \begin{tabular}{lcccccccccccc}
  \hline
  \hline
Object & R.A. (2000) & Dec. (2000) & $N_{\mathrm{OB}}$ & $\Sigma_{\mathrm{Peak,OB}}$& $\langle \mu \rangle$ & $\sigma_{\mu}$ & $\langle \Phi \rangle$ & $\sigma_{\Phi}$ & RV & $\sigma_{RV}$ &  $D$ & $\sigma_{V,2D}$  \\ 
          & (h m s) & ($^{\circ} \ ^{\prime} \ ^{\prime \prime}$) & & (stars pc$^{-2}$) & (mas yr$^{-1}$) &  (mas yr$^{-1}$) & ($^{\circ}$) & ($^{\circ}$) & (km s$^{-1}$) & (km s$^{-1}$) & (pc) & (km s$^{-1}$)  \\
  \hline
Cyg OB2 &&&&&&&&&&&& \\
Cluster A &20:32:27.22 & +41:25:36.4 & 33 & 1.85 & 4.86 & 0.23 & 211.4 & 3.2 & -10.1 & 4.6 & $6.2 \pm 0.4$ & $1.7 \pm 0.1$  \\ 
Cluster B & 20:33:18.02 & +41:17:44.9 & 39 & 2.95 & 5.16 & 0.24 & 211.7 & 1.9 & -11.3 & 6.2 & $6.3 \pm 0.4$ & $1.8 \pm 0.1$  \\
Halo       & 20:32:59.17 &  +41:23:44.9 & 59 & 0.06 & 4.83 & 0.42 & 211.3 & 4.0 & -11.5 & 4.2 & $28.7 \pm 1.8$ & $3.1 \pm 0.2$  \\
\hline
Car OB1 &&&&&&&&&&&& \\
Tr14 &10:43:57.59 & -59:33:28.1 & 43 & 3.22 & 6.87 & 0.21 & 287.1 & 2.1 & -6.2 & 4.8 & $5.1 \pm 0.5$ & $2.2 \pm 0.2$ \\ 
Tr15 &10:44:42.31 & -59:21:53.5& 14  & 1.16 & 6.42 & 0.19 & 287.8 & 1.5 &         &       & $5.8 \pm 0.5$ & $2.1 \pm 0.2$ \\
Tr16 &10:45:05.84 & -59:42:34.0 & 67 & 3.40 & 7.44 & 0.22 & 290.3 & 1.3 & -6.6 & 3.9 & $7.9 \pm 0.7$ & $2.4 \pm 0.2$  \\
Halo & 10:44:30.96  &   -59:50:46.7   & 88 & 0.03 & 7.09 & 0.65 & 288.7 & 3.5 &  -6.4 & 6.6 & $68.9 \pm 6.0$ & $7.1 \pm 0.6$ \\
\hline
\end{tabular}
\\
\begin{tabular}{@{}l@{}}
Col. (1) : Names of objects. Cols. (2) and (3) : The equatorial coordinates of stars. Col. (4) : The number of members. Col. (5) : Peak number density of OB \\ 
stars. Cols. (6) and (7) : Mean amplitude of proper motions and its dispersion. Cols. (8) and (9) : Mean orientation of proper motions and its dispersion. Cols. \\ 
(10) and (11) : Mean radial velocities and its dispersion. Col. (12) : Maximum diameter. Col. (13) : Two-dimensional 
velocity dispersion.\\
\end{tabular}
\end{table}
\end{landscape}

%%%%%%%%%%%%%%%%%%%%%%%%%%%%%%%%%%%%%%%%%%%%%%%%%%

%%%%%%%%%%%%%%%%% APPENDICES %%%%%%%%%%%%%%%%%%%%%

%\appendix



%%%%%%%%%%%%%%%%%%%%%%%%%%%%%%%%%%%%%%%%%%%%%%%%%%

% Don't change these lines
\bsp	% typesetting comment
\label{lastpage}
\end{document}